\documentclass{aa}  

\usepackage{graphicx}

\usepackage{txfonts}
\usepackage{textcomp}

\usepackage{longtable}
\usepackage{pdflscape}
\usepackage{natbib}
\usepackage{amsmath}
\usepackage{mathabx}

\usepackage{xcolor}

\usepackage[pdfstartview=FitH,      
            pdfborder={0 0 0},
            colorlinks=true,
            linkcolor=blue,
            citecolor=blue,
            urlcolor=blue,
            breaklinks=true,       
            bookmarksopen=true,     
            bookmarksnumbered=true  
            ]{hyperref}

\newcommand{\Kepler}{\textit{Kepler} }

\makeatletter
\renewcommand*\aa@pageof{, page \thepage{} of \pageref*{LastPage}}
\makeatother

\begin{document}

    \title{\Kepler Object of Interest Network\thanks{Ground-based photometry is only available 
    at the CDS via anonymous ftp to \url{cdsarc.u-strasbg.fr} (\url{130.79.128.5}) or via 
    \url{http://cdsarc.u-strasbg.fr/viz-bin/qcat?J/A+A}}}
    \subtitle{III. Kepler-82f: A new non-transiting $21 M_\Earth$ planet from photodynamical modelling}

    \author{J.~Freudenthal\inst{\ref{inst1}} \and 
    C.~von~Essen\inst{\ref{inst2},\ref{inst1}} \and
    A.~Ofir\inst{\ref{inst8}} \and
    S.~Dreizler\inst{\ref{inst1}} \and 
    E.~Agol\inst{\ref{inst5},\ref{inst6},\ref{inst7}}\and 
    S.~Wedemeyer\inst{\ref{inst3},\ref{inst4}} \and
    B.~M.~Morris\inst{\ref{inst5}} \and
    A.~C.~Becker\inst{\ref{inst5},\ref{inst17}} \and
    H.~J.~Deeg\inst{\ref{inst9}, \ref{inst10}} \and
    S.~Hoyer\inst{\ref{inst9}} \and
    M.~Mallonn\inst{\ref{inst12}} \and
    K.~Poppenhaeger\inst{\ref{inst12},\ref{inst13}} \and
    E. Herrero\inst{\ref{inst14},\ref{inst15}} \and
    I.~Ribas\inst{\ref{inst14},\ref{inst15}} \and
    P.~Boumis\inst{\ref{inst16}} \and
    A.~Liakos\inst{\ref{inst16}}
    }
    \institute{Institut f\"{u}r Astrophysik, Georg-August-Universit\"{a}t G\"{o}ttingen, Friedrich-Hund-Platz\,1, 37077 G\"{o}ttingen, Germany\label{inst1} \\ email:~\href{mailto:jfreude@astro.physik.uni-goettingen.de}{jfreude@astro.physik.uni-goettingen.de} \and 
    Stellar Astrophysics Centre, Aarhus University, Ny Munkegade 120, 8000 Aarhus, Denmark\label{inst2} \and
    Department of Earth and Planetary Sciences, Weizmann Institute of Science, Rehovot, 76100, Israel\label{inst8} \and
    Astronomy Department, University of Washington, Seattle, WA 98195, USA\label{inst5} \and
    Guggenheim Fellow\label{inst6} \and
    Virtual Planetary Laboratory, University of Washington, Seattle, WA 98195, USA\label{inst7} \and
    Rosseland Centre for Solar Physics, University of Oslo, P.O. Box 1029 Blindern, N-0315 Oslo, Norway\label{inst3} \and
    Institute of Theoretical Astrophysics, University of Oslo, P.O. Box 1029 Blindern, N-0315 Oslo, Norway\label{inst4} \and
    Amazon Web Services, Seattle, WA, 98121, USA \label{inst17} \and
    Aix Marseille Univ, CNRS, CNES, LAM, Marseille, France\label{inst9} \and
    Universidad de La Laguna, Dept. de Astrof\'\i sica, E-38206 La Laguna, Tenerife, Spain\label{inst10} \and
    Leibniz-Institut f\"{u}r Astrophysik Potsdam, An der Sternwarte 16, D-14482 Potsdam, Germany\label{inst12} \and
    Astrophysics Research Centre, Queen's University Belfast, Belfast BT7 1NN, UK\label{inst13} \and
    Institut de Ciències de l'Espai (IEEC-CSIC), C/Can Magrans, s/n, Campus UAB, 08193 Bellaterra, Spain\label{inst14} \and
    Institut d’Estudis Espacials de Catalunya (IEEC), Gran Capit\`a, 2-4, Edif. Nexus, 08034 Barcelona, Spain\label{inst15} \and
    Institute for Astronomy, Astrophysics, Space Applications and Remote Sensing, National Observatory of Athens, Metaxa \& Vas. Pavlou St., Penteli, Athens, Greece\label{inst16}
    }
   \date{Received Month DD, YYYY; accepted Month DD, YYYY}

   \abstract
   {The \Kepler Object of Interest Network (KOINet) is a multi-site network of telescopes around the globe organised for follow-up observations of transiting planet candidate \Kepler objects of interest (KOIs) with large transit timing variations (TTVs). The main goal of KOINet is the completion of their TTV curves as the \Kepler telescope stopped observing the original \Kepler field in 2013.}
   {We ensure a comprehensive characterisation of the investigated systems by analysing Kepler data combined with new ground-based transit data using a photodynamical model. This method is applied to the Kepler-82 system leading to its first dynamic analysis.}
   {In order to provide a coherent description of all observations simultaneously, we combine the numerical integration of the gravitational dynamics of a system over the time span of observations with a transit light curve model. To explore the model parameter space, this photodynamical model is coupled with a Markov chain Monte Carlo algorithm.}
   {The \object{Kepler-82b}/c system shows sinusoidal TTVs due to their near 2:1 resonance dynamical interaction. An additional chopping effect in the TTVs of Kepler-82c hints to a further planet near the 3:2 or 3:1 resonance. We photodynamically analysed \Kepler long- and short-cadence data and three new transit observations obtained by KOINet between 2014 and 2018. Our result reveals a non-transiting outer planet with a mass of $m_f=20.9\pm1.0\;M_\Earth$ near the 3:2 resonance to the outermost known planet, \object{Kepler-82c}. Furthermore, we determined the densities of planets b and c to the significantly more precise values $\rho_b=0.98_{-0.14}^{+0.10}\;\text{g cm}^{-3}$ and $\rho_c=0.494_{-0.077}^{+0.066}\;\text{g cm}^{-3}$.}
   {}

   \keywords{planets and satellites: dynamical evolution and stability – planets and satellites: detection – methods: data analysis – techniques: photometric – stars: individual: \object{Kepler-82} – stars: fundamental parameters}

   \maketitle

\section{Introduction}
There is no doubt about the impact that the \Kepler Space Telescope has had on the exoplanetary field. Among many other outstanding and benchmark contributions, such as the first possibly habitable planet with known radius \citep{Borucki2012}, and the first exoplanet ever found with two suns in its sky \citep{Doyle2011}, \Kepler data have allowed us to characterise planetary masses via transit timing variations \citep[TTVs, see e.g.][]{Fabrycky2012,Mazeh2013,Steffen2013}. Nonetheless, after four years of continuous monitoring of the same field of view, the nominal observations of Kepler came to an end. This left several \Kepler objects of interest (KOIs) without a proper characterisation, even though they presented large amplitude TTVs in the \Kepler data alone. To continue with the successful characterisation of planetary masses of KOIs via TTVs, we have organised the \Kepler Object of Interest Network\footnote{\url{koinet.astro.physik.uni-goettingen.de}} (KOINet). To date, results of our network comprise KOINet's first light \citep{vonEssen2017}, and the in-depth photodynamical characterisation of Kepler-9b/c \citep{Freudenthal2018}. While in the former we demonstrated KOINet's strategy and functionality, along with initial results on four KOIs, in the latter we were able to determine values for the planetary densities that are the most precise measurements in the regime of Neptune-like exoplanets. Furthermore, we predicted that the transits of Kepler-9c would disappear in about 30~years. These results arose from the combination of the \Kepler long- and short-cadence data with KOINet follow-up transit observations, along with a comprehensive and coherent analysis carried out with our photodynamical modelling. Similar analyses have likewise revealed precise planetary densities for other systems, like Kepler-117 by \cite{Almenara2015}, K2-19 by \citet{Barros2015}, WASP-47 by \citet{Almenara2016}, Kepler-138 by \citet{Almenara2018a}, and Kepler-419 by \citet{Almenara2018b}. In many of these cases the authors also demonstrated consistent planetary mass determinations from TTV and radial velocity (RV) measurements.

From amongst our KOINet targets we pinpointed Kepler-82 \mbox{(KOI 0880)} as an interesting system that deserves a detailed photodynamical analysis. The Kepler-82 system contains a total of four confirmed transiting planets. The two inner planets have periods of $P_d=2.38\;$d and $P_e=5.90\;$d, which were confirmed by \citet{Rowe2014}. The two outer planets have a period ratio close to 2:1 with $P_b=26.44\;$d and $P_c=51.54\;$d. This commensurability of the periods results in strong TTVs (see Fig. \ref{fig:ttv}), which led to the confirmation of the two outer planets a year before the inner planets \citep{Xie2013}. The inner two planets are not much affected by this dynamical interaction and also show no measurable dynamical interaction with one another. Yet Kepler-82e shows TTVs with an amplitude of about 15\;min, where the uncertainties of the transit times are of the same order, and the variations are without significant periodicity \citep{Holczer2016}. \citet{Ofir2018} found TTVs in Kepler-82d with an amplitude of $10.3_{-1.4}^{+1.8}\;$min and a frequency peak that just surpassed their significance criteria. The peak does not correspond to any expected dynamical frequency.

The first characterisation of the Kepler-82b/c TTVs was carried out by \citet{Xie2013}. The author found the TTVs to be sinusoidal as expected for near 2:1 mean-motion resonance (MMR) systems. In contrast with many other similar systems, the sinusoidal-shaped TTVs of both planets are not anti-correlated; instead the phase difference is close to zero. The author calculated the nominal masses assuming a two-planet system and found a relatively large mass ratio of $m_b/m_c\sim 10^{0.6}\sim4$, which means a very large density ratio of $\rho_b/\rho_c \sim 4\times(5.35/4)^3\sim 10$.  Another nominal mass computation by \citet{HaddenLithwick2014} indicates a smaller mass ($\sim3$) and density ratio ($\sim 7$).

A further characterisation was done by \citet{Ofir2018} by analysing periodograms of the TTVs of Kepler-82b/c. They found the most significant peak in the periodogram of Kepler-82b fits the 2:1 MMR super frequency. However, the highest amplitude peak of Kepler-82c is notably offset from the highest peak of Kepler-82b and the 2:1 MMR super frequency. Additionally, they found one other significant peak for Kepler-82b and three in Kepler-82c.

The following work includes the first dynamical analysis of the Kepler-82b/c system. We applied a photodynamical model to \Kepler data and ground-based follow-up observations from KOINet. With this we were able to constrain the planetary masses more precisely, and by including another non-transiting planet, most of the frequency peaks, can be explained. Furthermore, we were able to determine the stellar mass, radius and age from our results by combining the modelled stellar densities with spectroscopic values and comparing these values with stellar evolution models. 

The paper is structured as follows. The data acquisition and treatment within the KOINet is described in Sect.~\ref{sec:KOINet}. We present our own implementation of a photodynamical model in Sect.~\ref{sec:photdyn}. The detection of a third dynamically important non-transiting planet in the TTVs of Kepler-82c is described in detail in Sect.~\ref{sec:analysis}. The results from the analysis are discussed in Sect.~\ref{sec:Discussion}. We end the paper with a conclusion in Sect.~\ref{sec:conclusion}.

\section{KOINet data}
\label{sec:KOINet}
\begin{table*}
\caption{Characteristics of collected ground-based transit light curves of Kepler-82b/c, collected through KOINet.}  
\label{table:obscond}
\centering
\begin{tabular}{c c c c c c c c}
\hline\hline
Date        &   Planet  &   Telescope   &   $\sigma_\text{res}$ &   $N$   &   CAD     &   $T_\text{tot}$  &   TC  \\
yyyy.mm.dd  &           &               &   [ppt]               &       &   [sec]   &   [hours]         &       \\
\hline
2014.09.04  &   b       &   ARC 3.5 m    &   1.1                 &   235 &    60     &   4.6             &   - - B E O   \\
2017.07.08  &   c       &   NOT 2.5 m    &   1.4                 &   111 &    193    &   4.2             &   O I - - -   \\
2018.07.05  &   c       &   NOT 2.5 m    &   1.2                 &   112 &    179    &   5.7             &   O I - - -   \\
2015.05.26  &   b       &   SAO 6.5 m    &   0.3                 &   32  &   68      &   1.1             & Only off-transit data \\
2015.05.28  &   c       &   CAHA 2.2 m   &   10.8                &   75  &   86      &   1.9             & Only off-transit data \\
2015.07.19  &   c       &   CAHA 3.5 m   &   4.4                 &   292 &   102     &   6.8             & Parabolic solution was chosen \\
            &   c       &   LIV 2 m      &   1.5                 &   112 &   90      &   3.0             & Parabolic solution was chosen \\
            &   c       &   TJO 0.8 m    &   4.4                 &   163 &   69      &   4.4             & Parabolic solution was chosen \\
2018.07.05  &   c       &   KRYO 1.2 m   &   3.7                 &   242 &   83      &   5.8             & Only off-transit data \\
\hline
2014.07.23  &           &   IAC 0.8 m    & & & & & Corrupted data \\
            &           &   OLT 1.2 m    & & & & & Corrupted data \\ 
\hline
\end{tabular}
\tablefoot{ From left to right: the date on which the observations were carried out, in years, months and days; the planet the transit belongs to; an acronym for the telescope used to perform the observations; the precision of the data in parts-per-thousand (ppt), $\sigma_\text{res}$; the number of frames acquired during the night, $N$; the cadence of the data considering the readout time in seconds, CAD; the total duration of the observations in hours, $T_\text{tot}$; the transit coverage, TC. The letter code to specify the transit coverage during each observation is the following: O: out of transit, before ingress. I: ingress. B: flat bottom. E: egress. O: out of transit, after 
egress.}
\end{table*}

In order to organise the KOINet observations we calculated transit time predictions from the \Kepler observations as described in Sect.~2.5 of \citet{vonEssen2017}. In the case of Kepler-82b, a linear plus sine function was fitted to predict future times of transit. For Kepler-82c we provided two different predictions. One coming from a sine plus linear fit, and one from fitting a parabolic function as a turnover to the sine curve was not measured by the Kepler observations. The low precision in the transit time predictions of Kepler-82c in particular led to only a small fraction of KOINet Kepler-82 light curves with transits included. Between 2014 and 2018 eleven light curves of Kepler-82 were obtained, while only three of them show transit signals of Kepler-82b/c. 

Table~\ref{table:obscond} lists the main characteristics of the data presented in this paper, such as the observing telescope and the observation dates the precision of the data, the total duration of the observation, and the transit coverage. To increase the photometric precision of the collected data, we have, when possible, slightly defocused the telescopes \citep{Kjeldsen1992, Southworth2009}. Below is a brief description of the main characteristics of each of the telescopes involved in this work.

The Apache Point Observatory hosts the Astrophysical Research Consortium 3.5~m telescope (henceforth “ARC 3.5~m”), and is located in New Mexico, United States of America. The photodynamical analysis of Kepler-82 presented here includes one light curve taken with the ARC 3.5~m during our first observing campaign in 2014.

The 2.5~m Nordic Optical Telescope (NOT~2.5~m) is located at the Observatorio Roque de los Muchachos in La Palma, Spain. Currently, telescope time for KOINet is assigned via a large (three-years) program. Here, we present two light curves taken between the fourth and fifth observing seasons.

The 80 centimetre telescope of the Instituto de Astrofísica de Canarias (IAC~0.8~m) is located at the Observatorio del Teide, in the Canary Islands, Spain. The one transit light curve obtained in the first season of KOINet suffered from technical difficulties during the night. For this reason the resulting science frames were corrupted and, thus, it was impossible for us to properly reduce them.

The Oskar Lühning Telescope (OLT~1.2~m) has a 1.2~m aperture diameter and is located at the Hamburger Observatory in Hamburg, Germany. Kepler-82 was observed for one night in the first season of KOINet with OLT 1.2m. Unfortunately, the observation taken in 2014 suffered from technical difficulties. 

The Telescopi \textit{Joan Oró} is a fully robotic 80 centimetre telescope (TJO~0.8~m) located at the Observatori Astronomic del Montsec, in the north-east of Spain. The parabolic prediction of Kepler-82c was chosen as transit time for an observation. The obtained observation contains only off-transit data.

The fully robotic 2~m Liverpool telescope \citep[LIV~2~m;][]{2004SPIE.5489..679S} is located at the Observatorio Roque de los Muchachos and is owned and operated by Liverpool John Moores University. During the second season of KOINet a transit time predicted from parabolic TTVs was chosen for an observation. The resulting light curve does not contain a transit.

The Centro Astronómico Hispano-Alemán hosts, among others, a 2.2~m and a 3.5~m telescope (“CAHA~2.2~m” and “CAHA~3.5~m”). An observation was taken with each telescope. No transit is present in the light curves.

The MMT observatory, a joint venture of the Smithsonian Institution and the University of Arizona, is located on the summit of Mt. Hopkins in south-eastern Arizona, United States of America. The telescope has a collecting area of 6.5~m (SAO~6.5~m). The data collected with this telescope were of sub-millimagnitude precision, but taken outside transit due to bad scheduling decisions.

The National Observatory of Athens hosts the 1.2~m Cassegrain telescope of the Astronomical Station Kryoneri (KRYO~1.2~m). For the last 40 years the telescope has been operational, with an extensive upgrade taking place in 2016. Data collected with this telescope were of good quality, however taken outside transit.

All collected observations underwent the KOINet reduction pipeline, and a preliminary analysis for deriving reliable errorbars and the detrending components. This process is described in \citet{vonEssen2017} and \citet{Freudenthal2018}.

\section{The photodynamical model}
\label{sec:photdyn}
For the KOINet data analysis we developed a simultaneous transit light curve model for all observations of each system that takes the system dynamics into account. This allows us to determine the planetary masses in addition to the transit parameters. A full description of our photodynamical model can be found in \citet{Freudenthal2018}. Briefly, we combine a numerical integration of the whole system over the time span of observations, and from the output sky positions (projected distance of each planet to the star) we calculate the transit light curve. We use a second-order mixed-variable symplectic (MVS) algorithm  to perform the numerical integration as implemented in our python-wrapper for \texttt{mercury6} \citep{Chambers1999}. The integrator is complemented by first-order post-Newtonian correction \citep{1995PhRvD..52..821K}, and we correct the individual times for the light-travel-time effect for each planet. From the numerical integration of the system we extract the planet-to-star centre distances to calculate the light curve through the transit model of \citet{MandelAgol2002}. Here we use the \texttt{occultquad} routine with the quadratic limb-darkening law implemented.

As in \citet{Freudenthal2018}, the numerical integration is done on a coarse grid, and only in the vicinity of transits is the integration refined with a time step of 0.01\;d. The coarse grid is optimised to give the shortest possible computation time with sufficient accuracy. For this system a time step of a hundred-twentieth of the period of the innermost included planet was used. For long-cadence data we take the finite integration time into account\citep{Kipping2010}. Hence, we compute the transit light curve with a time step of $\sim1\;$minute and rebin it to the cadence of the data points.

Our photodynamical model is coupled to the Markov chain Monte Carlo (MCMC) \texttt{emcee3} algorithm \citep{emcee}. All fitting parameters have uniform priors with broad boundaries chosen to avoid non-physical results. A detailed description of the model parameters can be found in \citet{Freudenthal2018}. To summarise, the model requires the mass, $m$, and the radius $R$ of the central star, as well as the two quadratic limb darkening coefficients, $c_1$ and $c_2$, that reflect the wavelength response of the optical setup of each telescope per instrument, and per planet, $p$ ($p\in \{b,c,f\}$ from Sect.~\ref{sec:analysis} and for example in the Tables~\ref{table:results1} and \ref{table:results}) the parameters are described below.

A mass ratio is needed. For the innermost planet the ratio to the central star, $m_1/m_S$, is taken and for all other planets the ratio to the next inner one, $m_{p}/m_{p'}$. Secondly, a parameter to calculate the semi-major axis, $a$, is needed. In the case of transiting planets it is calculated from the mean period, $P_p$ and as a free parameter a correction factor, $a_{p,\text{corr}}$:
$$a_p=\left(\frac{ P_p^2 G (m_S+m_p)}{4\pi^2}\right)^{1/3}\cdot a_{p,\text{corr}}\;,$$
with the gravitational constant, $G$. We fitted a linear ephemeris $T= \Delta T_{p,0}+P_p\cdot n$ to the transit times, $T$, giving us the mean period $P_p$ and an offset $\Delta T_{p,0}$. For non-transiting planets the semi-major axis is calculated from the period given by a period ratio to the next inner planet. Furthermore, the eccentricity, $e_p$, is needed. The orbital angles, inclination, $i_p$, argument of the periastron, $\omega_p$, and the longitude of the ascending node, $\Omega_p$, are needed. Whereas the latter is fixed to zero for the innermost planet, the other values are given relative to the innermost planet. The instantaneous position of the planets at a given reference time needs to be defined. We take the mean anomaly, $M_p$, as measurement for the position of each planet. This angle is calculated from the mean period, $P_p$, as well as the offset, $\Delta T_{p,0}$. As a free parameter, we have an addition to this derived mean anomaly, $M_{p,\text{corr}}$:
$$M_p=M_{p,\text{Kepler}}-\frac{2\pi}{P_p}\Delta T_{p,0} + M_{p,\text{corr}}$$
with the mean anomaly at transit time calculated for a Kepler orbit from the argument of periastron and eccentricity, $M_{p,\text{Kepler}}$, and the second term is giving the difference between the mean anomaly at transit time and the mean anomaly at the starting time of the integration. That means the free parameter $M_{p,\text{corr}}$ is giving the correction from a pure Keplerian orbit due to the interaction with the other planets. Lastly, The planet-to-star radius ratio, $R_p/R_S$, only for transiting planets needs to be given.

We treated \Kepler data and ground based observations of KOINet as the description in \citet{Freudenthal2018}. From \Kepler photometry we extracted the transit duration symmetrically around each transit mid point four times. To account for intrinsic stellar photometric variability we normalised each transit light curve dividing it by a time dependent second-order polynomial optimised on the off-transit data points. The coefficients of this parabola are derived through a simple least-squares minimisation routine. As previously mentioned, for long-cadence data, the photodynamical light curve model is oversampled by a factor of 30 and rebinned to the actual data points. This procedure is not necessary for short-cadence data. The high signal-to-noise ratio (S/N) of Kepler data allows us to include the quadratic limb darkening coefficients into our free parameters set. This allows for a more realistic inclination and star and planetary radii determination due to the good constrained transit shape.

Due to the lower S/N of the ground-based data, we fixed the quadratic limb darkening coefficients to values which are derived as described in \citet{vonEssen2013} from stellar parameters for the Johnson-Cousins R-band filter, which we used for all of our observations. For stellar parameters closely matching the ones of Kepler-82 \citep{Petigura2017}, the derived limb darkening coefficients are $c_1 = 0.52$ and $c_2 = 0.14$. The best-matching coefficients of the detrending components, derived during the first data analysis (in Sect. \ref{sec:KOINet}), for each ground-based observation are calculated as a linear combination at each call of the photodynamical model.

\begin{figure*}
\sidecaption
  \includegraphics[width=12cm]{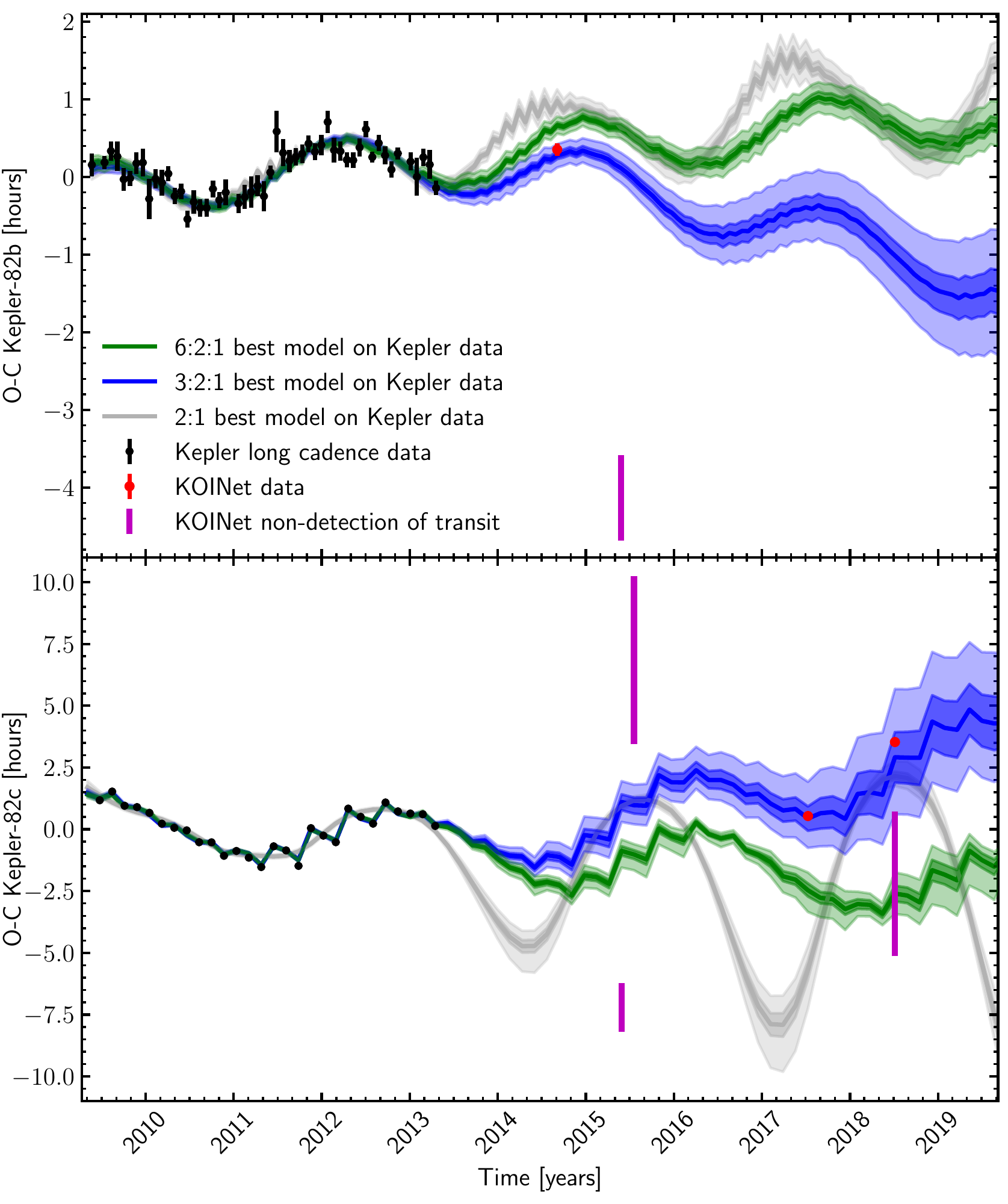}
     \caption{O-C diagrams of Kepler-82b at the top and Kepler-82c at the bottom with transit times from modelling the transits individually. The black points refer to the transit data from the Kepler telescope. The red points are the individual transit times from the new KOINet observations (plotted in Fig.~\ref{fig:NewTransits}). The violet lines show observed epochs but with non-detections of transit. The green area indicates the 99.7\;\% (light)  and the 68.3\;\% (dark) confidence interval of the 6:2:1 resonance model optimised on Kepler long- and short-cadence data only, derived from 1\;000 randomly chosen models out of the MCMC posterior distribution; the green line is the median. The blue areas indicate the 3:2:1 resonance model solution. The grey areas present the 2:1 resonance model.}
     \label{fig:ttv}
\end{figure*}
\begin{figure*}
\sidecaption
  \includegraphics[width=12cm]{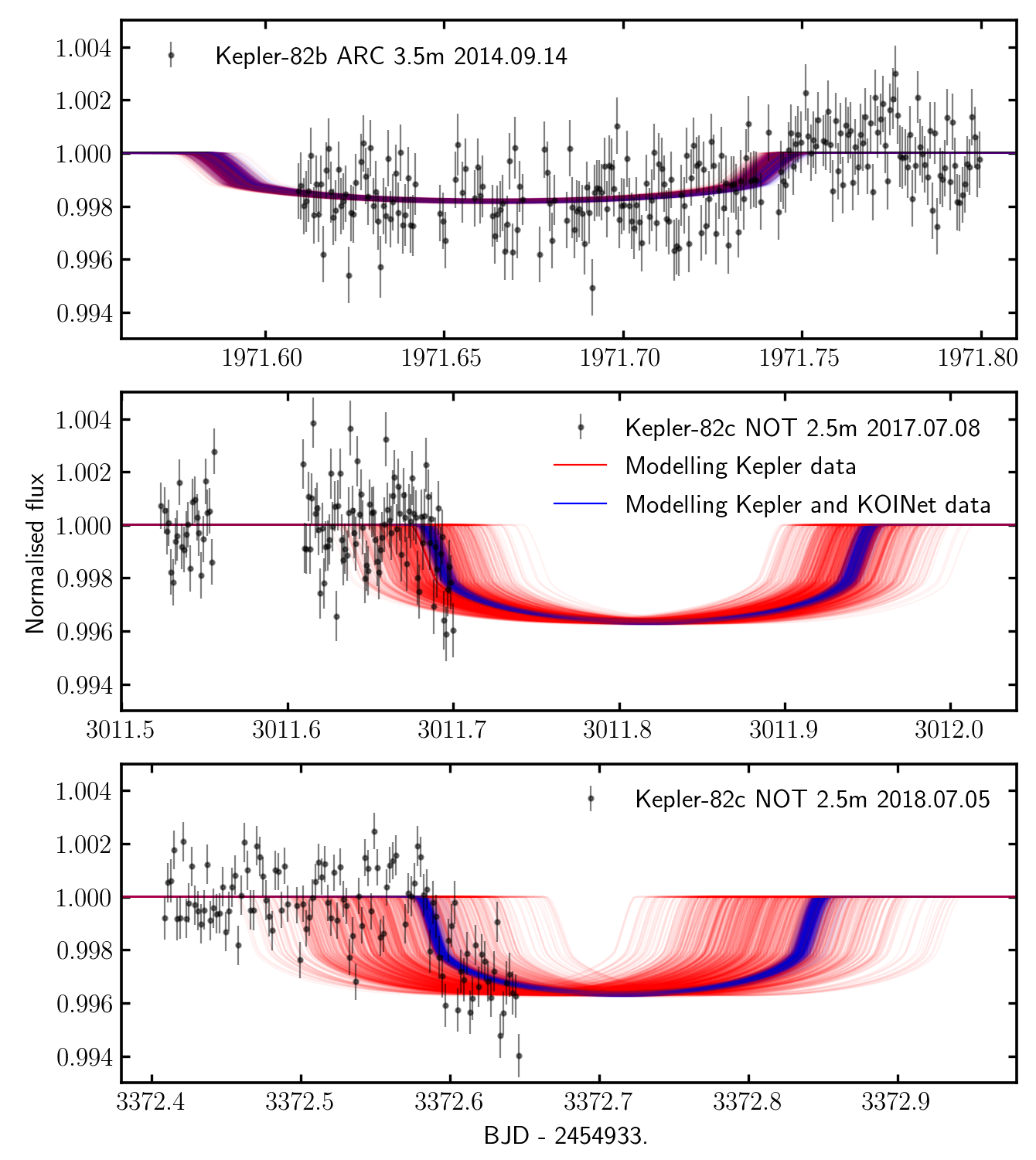}
  \caption{KOINet transit light curves of Kepler-82. The three transit light curves (black) are overplotted with 1000 random models of the 3:2:1 resonance optimised using \Kepler long- and short-cadence data (red) and including these KOINet observations (blue).}
\label{fig:NewTransits}
\end{figure*}

\section{Dynamical analysis of Kepler-82}
\label{sec:analysis}
\begin{table}
\caption{Planetary and Stellar parameters from photodynamical analysis of the 3:2:1 resonance model on \Kepler data and the three KOINet transit light curves.}
\label{table:results1}
\centering
\begin{tabular}{l c }
\hline\hline															
Parameter &  Values \\
\hline															
\noalign{\smallskip}							
\multicolumn{2}{l}{\bf Kepler-82b}							\\
$m_b/m_S$	& $	0.0000401	_{-	0.0000025	}^{+	0.0000028	}$ \\
$m_b^* \;[M_\Earth]$	& $	12.15	_{-	0.87	}^{+	0.96	}$ \\
$a_{b,\text{corr}}$	& $	0.9999606	_{-	0.0000041	}^{+	0.0000042	}$ \\
$a_b^* \;[\text{AU}]$	& $	0.1683			\pm	0.0020	$ \\
$P_b^* \;[\text{d}]$	& $	26.44			\pm	0.48	$ \\
$e_b$	& $	0.0033	_{-	0.0017	}^{+	0.0019	}$ \\
$i_b\;[^\degree]$	& $	89.052	_{-	0.096	}^{+	0.049	}$ \\
$\Omega_b\;[^\degree]$	& 	$0$ (fixed)					\\
$\omega_b\;[^\degree]$	& $	236	_{-	26	}^{+	23	}$ \\
$M_{b,\text{corr}}\;[^\degree]$	& $	-0.025	_{-	0.020	}^{+	0.019	}$ \\
$M_b^* \;[^\degree]$	& $	13	_{-	23	}^{+	26	}$ \\
$R_b/R_S $	& $	0.04159	_{-	0.00045	}^{+	0.00049	}$ \\
$R_b^* \;[R_\Earth]$	& $	4.07	_{-	0.10	}^{+	0.24	}$ \\
$\rho_b^* [\text{g\;cm}^{-3}]$	& $	0.98	_{-	0.16	}^{+	0.11	}$ \\
\noalign{\smallskip}							
\multicolumn{2}{l}{\bf Kepler-82c}							\\
$m_c/m_b$	& $	1.14	_{-	0.13	}^{+	0.14	}$ \\
$m_c^* \;[M_\Earth]$	& $	13.9	_{-	1.2	}^{+	1.3	}$ \\
$a_{c,\text{corr}}$	& $	0.999947	_{-	0.000019	}^{+	0.000018	}$ \\
$a_c^* \;[\text{AU}]$	& $	0.2626			\pm	0.0032	$ \\
$P_c^* \;[\text{d}]$	& $	51.54			\pm	0.94	$ \\
$e_c$	& $	0.0070	_{-	0.0018	}^{+	0.0016	}$ \\
$i_c\;[^\degree]$ config. I	& $	90.15	_{-	0.22	}^{+	0.18	}$ \\
$i_c\;[^\degree]$ config. II	& $	89.78	_{-	0.15	}^{+	0.17	}$ \\
$\Omega_c\;[^\degree]$	& $	1.6			\pm	2.1	$ \\
$\omega_c\;[^\degree]$	& $	162	_{-	20	}^{+	12	}$ \\
$M_{c,\text{corr}}\;[^\degree]$	& $	-0.507			\pm	0.020	$ \\
$M_c^* \;[^\degree]$	& $	131	_{-	12	}^{+	20	}$ \\
$R_c/R_S $	& $	0.05453	_{-	0.00053	}^{+	0.00068	}$ \\
$R_c^* \;[R_\Earth]$	& $	5.34	_{-	0.13	}^{+	0.32	}$ \\
$\rho_c^* [\text{g\;cm}^{-3}]$	& $	0.494	_{-	0.083	}^{+	0.070	}$ \\
\noalign{\smallskip}							
\multicolumn{2}{l}{\bf Kepler-82f}							\\
$m_f/m_c$	& $	1.50	_{-	0.13	}^{+	0.16	}$ \\
$m_f^* \;[M_\Earth]$	& $	20.9			\pm	1.0	$ \\
$P_f/P_c$	& $	1.46940	_{-	0.00022	}^{+	0.00023	}$ \\
$P_f^*$ [d]	& $	75.732			\pm	0.012	$ \\
$a_f^* \;[\text{AU}]$	& $	0.3395			\pm	0.0041	$ \\
$e_f$	& $	0.0014	_{-	0.0010	}^{+	0.0018	}$ \\
$i_f\;[^\degree]$ config. I	& $	86.30			\pm	0.56	$ \\
$i_f\;[^\degree]$ config. II	& $	93.62	_{-	0.72	}^{+	0.56	}$ \\
$\Omega_f\;[^\degree]$ 	& $	1.6	_{-	2.1	}^{+	2.2	}$ \\
$\omega_f\;[^\degree]$	& $	62	_{-	47	}^{+	70	}$ \\
$M_f \;[^\degree]$	& $	125	_{-	70	}^{+	47	}$ \\
\noalign{\smallskip}							
\multicolumn{2}{l}{\bf Kepler-82}							\\
$m_S [M_\Sun]$	&	$0.91\pm0.03$ \citep[fixed,][]{Johnson2017}					\\
$R_S\;[R_\sun]$	& $	0.898	_{-	0.018	}^{+	0.042	}$ \\
$\rho_S^* [\text{g\,cm}^{-3}]$	& $	1.77	_{-	0.23	}^{+	0.11	}$ \\
$c_{1,\text{\it Kepler}}$	& $	0.522	_{-	0.075	}^{+	0.054	}$ \\
$c_{2,\text{\it Kepler}}$	& $	0.12	_{-	0.09	}^{+	0.14	}$ \\
\hline
\end{tabular}
\tablefoot{Listed are the median values and $68.26$\% confidence interval from the MCMC posterior distribution. The osculating orbital elements are given at a reference time, BJD = 2454933.0. $^{(*)}$Derived, not fitted parameters.}
\end{table}
In the following sections we outline the detection of a fifth, non-transiting planet in the Kepler-82 system, which is required to explain the available data. We call the planet Kepler-82f hereafter.

In this work we analyse the transit light curves of the outer two planets of Kepler-82, b and c. These planets have a period ratio close to the 2:1 resonance. The inner two, d and e, show no strong TTV amplitudes and especially no frequencies due to interaction with the outer two \citep{Ofir2018}. In a first step we determined the transit times from long-cadence \Kepler data with the procedure described in Sect.~4.1 of \citet{vonEssen2017}. In addition to the near resonant interaction with Kepler-82b, the transit times of Kepler-82c show a strong 'chopping' component, which is visible by a sudden jump in the transit time following every three consecutive transits which show drifting transit times.  The period of chopping is controlled by the times between conjunctions of planet $c$ and the fifth planet, given by the synodic period
$$P_{\text{syn}}=\left|\frac{1}{P_\text{out}}-\frac{1}{P_\text{in}}\right|^{-1}.$$
Since the jump in chopping is seen every three transits of planet $c$, this indicates that the synodic period is either $3 \times P_c$ or $3/2 \times P_c$, which would give a dependency of the acceleration and the deceleration during the orbits of the inner planet from three times its period. These synodic periods can be created by an outer planet near the 3:2 or 3:1 resonance with planet c. Based on the synodic period of planet $c$, an inner planet near the 3:4 or 3:5 resonance would also be possible; however, such a planet would be near a 3:2 or 6:5 resonance with Kepler-82b, and would then induce a strong signal in its TTVs. Such a TTV signal is not measured; hence the fifth planet must orbit exterior to planet $c$. 

For this reason we optimised the parameters of the two outer unknown planet configurations (from now on the 3:2:1 and 6:2:1 resonance models, for convenience we skip the more accurate notation of the planets being near resonant) in a photodynamical model applied to the \Kepler long-cadence (quarters 1-6) and short-cadence (quarters 7-17) data. From the \Kepler data alone, both of the resonance models show the same probability. The prediction for the transit times, however, start to diverge rapidly after the \Kepler mission terminates, as visualised in Fig.~\ref{fig:ttv}. The figure shows the transit times with a linear ephemeris subtracted (observed minus calculated, thus henceforth, O-C diagram) of Kepler-82b at the top and of Kepler-82c at the bottom. For Kepler-82b the models start to differ within $3\sigma$ by the end of 2015 and for Kepler-82c by mid 2014. The three KOINet transit light curves (plotted in Fig.~\ref{fig:NewTransits}; in the O-C diagram the transit times are indicated in red) show a clear preference for the 3:2:1 resonance model. In addition, the latest KOINet observation where no transit is measured clearly contradicts the 6:2:1 resonance model prediction.

On this account we re-optimised the 3:2:1 resonance model parameters to the \Kepler data complemented by the three KOINet transit light curves. The resulting planetary and stellar parameters from this fit can be found in Table~\ref{table:results1}. In the appendix Table~\ref{table:results} lists the planetary and stellar parameters from all model optimisation done in this work. The tables shows from top to bottom the modelled and derived values of Kepler-82b, Kepler-82c, the new planet, Kepler-82f, and the central star. The osculating orbital elements are given at the reference time BJD~=~$2454933.0$, $100~$days later than the standard Kepler reference time (BKJD). 

For comparison we also optimised the transiting 2-planet system (2:1 resonance model) on the \Kepler long- and short-cadence data. The results are listed as well and presented in the O-C diagram (Fig.~\ref{fig:ttv}) as grey areas.

\begin{table*}
\caption{Properties of the optimisation of different models paired with different data sets.}
\label{table:modelling}
\centering
\begin{tabular}{l c c c c}
\hline\hline															
Parameter & Kepler data &   Kepler data &   Kepler data & Kepler \& \\
&   & & & KOINet data \\
&   2:1   &	6:2:1   &   3:2:1    &   3:2:1 	\\
\hline															
Walkers    &  56   &    59   &    91   &    86   \\
Iterations per walker  &    20~000   &  75~000   &  100~000   &  175~000   \\
Iterations burn-in per walker   &  10~000   &   25~000   &  25~000   & 25~000   \\
Autocorrelation length    & 1~136   &   11~498   &   14~595   &    22~422   \\
Independent samples in total &    986    &  385   &  624   &  671  \\
Degrees of freedom  &   63~372   &  63~365   &  63~365   & 63~823 \\
Best $\chi^2_\text{red}$ &  1.239   &   1.225   &  1.224   &  1.227   \\
\hline
\end{tabular}
\end{table*}

\subsection{Details of optimisation}
We initially optimised the different planetary system models (described later in this section) on the transit times, fixing all transit shape determining parameters to narrow the parameter space for the photodynamical analysis. We used the median values and the $3\sigma$ interval of this analysis for a Gaussian random choice of starting parameter sets. The parameters describing the transit shape -- the  inclination, limb darkening coefficients and planet and star radii -- are taken from the individual transit fits.

We fixed the stellar mass to its literature value of ${m_S=0.91 M_\Sun}$ \citep{Johnson2017} during the TTV and the photodynamical analysis. The uncertainty on the stellar mass, ${\sigma_{m_S}=0.03 M_\Sun}$, is applied to the derived parameters that depend on it via error propagation. In particular this affects the planetary masses, semi-major axes, and periods.

Optimising a linear ephemeris to the \Kepler transit times of Kepler-82b/c, we obtained the offsets $\Delta T_{b,0} = 41.23683$~d and $\Delta T_{c,0} = 22.52550$~d as intercepts, and the mean periods $P_b = 26.44404770$~d and $P_c = 51.53912652$~d as slopes. The offsets and mean periods are used for the determination of the semi-major axes and the mean anomalies, as described previously in Sect.~\ref{sec:photdyn}.

The properties of all of the photodynamical model optimisation procedures on the transit light curves are given in Table~\ref{table:modelling}. Listed are the parameters as follows.  In the first row the number of walkers used for extracting the final results are given. We initialised with more walkers; however, a variable number of walkers ended in higher $\chi^2$ minima. Next, the number of iterations we obtained per walker are given, followed by the number of iterations we used as initial burn-in. From the MCMC posterior distribution we calculated the autocorrelation length according to \citet{GoodmanWeare2010}, but averaging over the autocorrelation function per walker instead of averaging directly over the walker values, as discussed in the blog by Daniel Foreman-Mackey\footnote{\url{https://dfm.io/posts/autocorr/}}. The given autocorrelation length allows us to derive the effective number of individual samples. The last two rows contain the degree of freedom (dof) of the optimisation and the best reduced $\chi^2$ value. We note a significant deviation from one in the reduced $\chi^2$ values which is unexpected considering the high dof numbers. For this reason, we quadratically add a systematic error of $\sqrt{\chi_\text{red}^2}-1 \sim10\%$ to the model parameter uncertainties in Table~\ref{table:results1}~and~\ref{table:results}. 

\begin{figure}
\resizebox{\hsize}{!}{\includegraphics{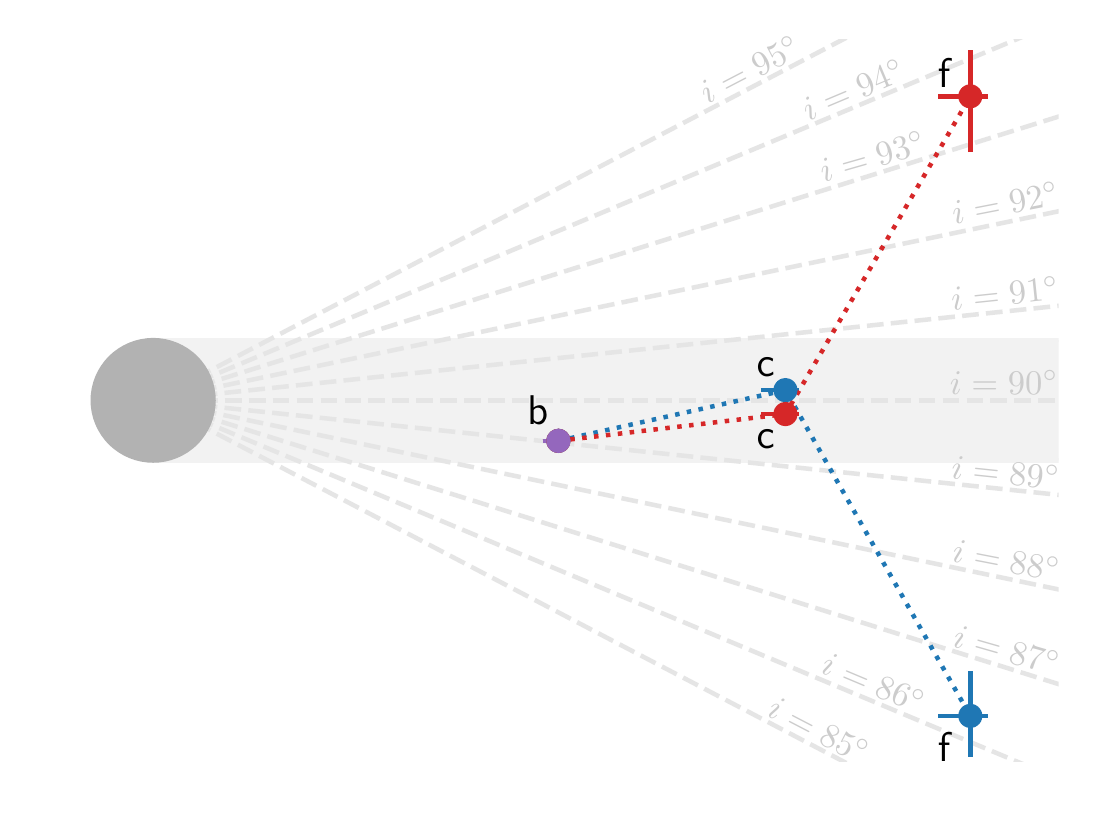}}
\caption{Configurations of the Kepler-82 system. With the star in grey on the left side and the observer on the right side this shows the two different configurations: b in violet has the same position in both, c and f in red shows configuration~I and in blue configuration~II. The grey area indicates the region of impact parameters below one. The distances are not true to scale with the stellar radius, therefore a few inclination values are indicated as dashed grey lines. A similar plot with true scales can be found in the appendix in Fig.~\ref{fig:config2}.}
\label{fig:config}
\end{figure}

While optimising the 3:2:1 resonance model we realised that we could actually derive the entire orbit of the non-transiting planet. By this we mean that we could constrain the inclination -- which avoids transit -- and the other orbital angles: the longitude of periastron, the longitude of  ascending node, and the mean anomaly. We found two different configurations with $i_b$ constrained to below $90^\degree$. The first has $i_c<90^\degree$ and $i_f>90^\degree$ (henceforth configuration~I), the second is the opposite with $i_c>90^\degree$ and $i_f<90^\degree$ (configuration~II). The configurations are visualised in Fig.~\ref{fig:config}, where the impact parameter of the planets is plotted against the distance to the star. The values and uncertainties are derived from 1~000 randomly chosen results from modelling the \Kepler and KOINet data in configuration~I in red and in configuration~II in blue. Both configurations have the same probability and are equivalent in all other parameters. This means that the transit time predictions and shape are the same for both configurations. The other two configurations, with either both planets having inclinations below $90^\degree$ or both above $90^\degree$, are not chosen by the MCMC optimisation, although allowed and included in the starting positions of the walkers. We modelled both configurations individually with the same number of iterations and combined all resulting walkers to extract the results. Given that the KOINet transit times are located at the 3:2:1 resonance model predictions, we optimised this model on these light curves together with the \Kepler data again in both of the configurations. 

Following this detection we also set the inclination of the non-transiting planet in the 6:2:1 resonance model as a free parameter. In this case the inclination did not avoid the transit region, though it spans a large area where the majority of solutions is in the non-transiting region (about 88~\% in a conservative calculation of the impact parameter~$b$). Nonetheless, we inspected the \Kepler data for these transits. Based on the mass ratio to Kepler-82c of $23_{-1.9}^{+2.5}$ we can expect transits of larger depths compared with the other system's planets. Such transits are not detected. 

\subsection{Results}
Along with the optimised parameters listed in Table~\ref{table:results1}~and~\ref{table:results} we display the KOINet transit light curves in Fig.~\ref{fig:NewTransits} in black. These are overplotted with 1\;000 model solutions randomly chosen from the MCMC posterior distribution from analysing {\it only} \Kepler data in red, and including these KOINet transit light curves in blue with the 3:2:1 resonance model. Similar to the O-C plot in Fig.~\ref{fig:ttv} we show the TTV behaviour for the 3:2:1 resonance model optimised on all available transit light curves in comparison to the optimisation on \Kepler data only in the appendix in Fig.~\ref{fig:ttv2_1}. Including the KOINet transit observations led to a narrowing of the transit time predictions of Kepler-82c (visible in the Fig.~\ref{fig:NewTransits} and Fig.~\ref{fig:ttv2_1}) and the shrinkage of the mass uncertainties of Kepler-82b (see Table~\ref{table:results}). The transit time predictions for the next fifteen years are listed in Table~\ref{table:ttv}. Finally, the parameter correlations are visualised in a corner plot in Fig.~\ref{fig:corner}.

\section{Discussion}
\label{sec:Discussion}
The most prominent signal in the TTVs of the Kepler-82b/c system is the dynamical interaction with each other due to its near 2:1 resonance configuration. \citet{Xie2013} calculated nominal masses from the amplitudes of these TTVs and a derived stellar mass from $\log g$ and $R_S$ under the assumption of a 2\hbox{-}interacting\hbox{-}planet system. Their derived masses for Kepler-82b/c are $m_b = 87.0_{-22.4}^{+251.8}\;M_\Earth$ and $m_c = 19.1_{-4.9}^{+55.5}\;M_\Earth$, respectively. Additionally, they derive the planetary radii from single transit fitting to $R_b = 4.00\pm1.82\;R_\Earth$ and $R_c = 5.35\pm2.44\;R_\Earth$. With these values they propose a density ratio of $\sim 10$ for the planets. In an initial model we tested this 2-planet system with our photodynamical analysis. We found planetary masses and radii with much smaller uncertainties (see Table~\ref{table:results}) that agree within their errorbars with the values calculated by \citet{Xie2013}. The density ratio of our result is even higher with $\rho_b/\rho_c\sim14$.

The stellar parameters of this analysis show significant deviations from literature values that are derived by spectroscopic observations. The stellar radius with ${R_S=1.186_{-0.077}^{+0.074}}$ is more than $1\sigma$ higher than the measurement by \citet{Johnson2017} ($R_S=0.99_{-0.08}^{+0.10}\;R_\Sun$) and the quadratic limb darkening coefficients calculated by \citet{Claret2011} (ATLAS model) to $c_1=0.4695$ and $c_2=0.2240$ do not fall within the modelled values ($c_1=0.31_{-0.17}^{+0.20}$, $c_2=0.66_{-0.32}^{+0.24}$).

These stellar parameters as well as the planetary masses, and with these the densities, become more plausible in their values when including a third planet in the dynamical analysis. The signal of such a planet is clearly visible in the TTVs of Kepler-82c as a jump every three consecutive transits (see Fig.~\ref{fig:ttv}). As explained in Sect.~\ref{sec:photdyn}, two different configurations of a three-planet system can explain this chopping effect in the \Kepler data. Both of these include another outer non-transiting planet, near the 3:1 or 3:2 period resonance to Kepler-82c. Including either of these planets dramatically reduces the mass of Kepler-82b, and thus also reduces the ratio of the density Kepler-82b to c. Both system models are very similar in probability for Kepler data, the 6:2:1 resonance model has a slightly higher $\chi^2_\text{red}$ than the 3:2:1 resonance system. With KOINet data we were able to distinguish between these two models. The detected transits fall at the 3:2:1 model prediction, and one of the observations where no transit is observed precludes the 6:2:1 model predicted transit time. In the following we refer to the 3:2:1 resonance model solution on \Kepler and KOINet data when not differently specified. 

The density ratio of the resulting Kepler-82b/c planets reduces to a factor of $\sim2$. Such a ratio is no longer very unusual; the values are discussed in the context of the literature below by visualising them in a mass-radius diagram. The density of the new planet can not be determined as, due to the lack of transits, the radius is not measurable. In addition, the stellar radius and the limb darkening values fit in with the literature values within $1\sigma$\hbox{-}uncertainty.

At the same time the predicted RV signal reduces from an amplitude of $\sim50\;\text{m\,s}^{-1}$ for the 2-planets system to about $\sim7.5\;\text{m\,s}^{-1}$ for the 3-planets system near 3:2:1 resonance. With Kepler-82 being a relatively faint star (Kp$=15.158$), such a signal is not measurable with current instruments.

\subsection{Previously proposed planets}
\citet{Boivard2015} predicted two additional planets in the Kepler-82 system with periods of $11.8\pm2.0$~days and $120\pm20$~days based on the Titius-Bode relation. Neither the new planet proposed here near the 3:2 resonance to Kepler-82c, nor the less viable option with a planet near the 3:1 resonance, matches the position of one of the predicted planets. The predicted outer planet is in between the two possibilities within $3\sigma$ distance to each of them.

\subsection{Dynamical stability}
Subsequent to the photodynamical analysis we tested the dynamical stability of the modelled systems. With the same integrator, the second-order mixed-variable symplectic algorithm implemented in the \texttt{mercury6} package by \citet{Chambers1999}, we extend the numerical simulation of the best found solution for each system configuration to 10~Gyr. For this application the post-Newtonian correction \citep{1995PhRvD..52..821K} was implemented as well. The integration is done with a time step size of 1~day which is roughly a twentieth of the innermost planet considered in our analysis (Kepler-82b). This gives a good compromise between a sufficient sampling for small integration errors and a reasonable computation time. We tested the stability of the 2:1 resonance 2~planets solution, the 6:2:1 resonance system as well as the 3:2:1~resonance 3~planets model that is preferred by the KOINet data. All of these system configurations survived the 10~Gyr integration; only the 2-planet system showed chaotic parameter evolution. 
A closer inspection of the 6:2:1 resonance system long-term behaviour showed that given this model we are observing the transiting planets b and c at a minimum in periodically changing eccentricities. The values are ranging in roughly $e_b=0.002 - 0.08$ and $e_c=0.004 - 0.06$. The probability for the planets to be in this minimum at observation time is below 10\%, making this scenario even less likely.

Another indication for stability is the planets to be near resonant, but not in resonance. We checked that the modelled planets are not in resonance through calculating the resonant angles \citep{Morbidelli2002}, as well as the Laplace resonant angle. All angles are circulating and do not librate, which would be the sign for the planets to be in resonance.

\subsection{TTV frequencies}
\begin{table}
\caption{Comparison between TTV inducing frequencies calculated from periods of the system solution to measured TTV frequencies by \citet{Ofir2018}. Super, chopping, and orbital frequencies are given; from left to right the table shows the considered planets, the computed frequency, a match in the \citet{Ofir2018} results (if any), and the matching frequency.}
\label{table:freq}
\centering
\begin{tabular}{l r l r}
\hline\hline															
Planets & $f_\text{calc}$ [$10^{-4}$d$^{-1}$] & TTV matching & $f_\text{Ofir}$ [$10^{-4}$d$^{-1}$] \\
\hline
\multicolumn{4}{l}{Super frequencies:}	\\
b and c & $9.86 \pm 0.22$ & main peak in b & $9.82_{-0.45}^{+0.39}$\\
b and f & $18.06 \pm 0.33$  & one peak in c & $17.9\pm3.2$ \\
c and f & $8.18 \pm 0.26$ & main peak in c & $8.15\pm0.12$ \\
\multicolumn{4}{l}{Chopping frequencies:}  \\
b and c & $184.2\pm 3.1$  & no matching peak &  \\
b and f & $246.1 \pm 4.1$   & no matching peak & \\
c and f & $61.9 \pm 1.0$   & one peak in c & $58.9\pm3.2$ \\
\multicolumn{4}{l}{Orbital frequencies:}	\\
b & $378.2 \pm 6.3$ & no matching peak &   \\
c & $194.0 \pm 3.2$ & no matching peak &   \\
f & $132.1 \pm 2.2$ & no matching peak &   \\
\hline
\end{tabular}
\end{table}
The frequencies in the TTVs of the Kepler-82 system were analysed by \citet{Ofir2018}. In the TTVs of Kepler-82b they found, besides the main peak at $9.82_{-0.45}^{+0.39}\cdot 10^{-4}\;\text{d}^{-1}$, another significant frequency peak at $(101.5\pm2.8)\cdot 10^{-4}\;\text{d}^{-1}$. The main frequency peak of Kepler-82c is at $(8.15\pm0.12)\cdot 10^{-4}\;\text{d}^{-1}$. In addition to that they found three more peaks in the TTVs at ${((17.9,\;58.9, \;68.9)\pm3.2)\cdot 10^{-4}\;\text{d}^{-1}}$. Except for the main peak of Kepler-82b belonging to the super frequency of the near 2:1 resonance with Kepler-82c, they could not explain the detected frequencies with super frequencies of all of the mean motion resonances, orbital frequencies, chopping frequencies, or stroboscopic frequencies of the confirmed planets in the system. 

In the same manner we computed the super frequencies of all mean motion resonances, orbital frequencies, and chopping frequencies of our resulting system from photodynamical analysis. The calculated frequencies are listed in Table~\ref{table:freq}. With the exception of two measured frequencies, we can explain them with interactions of the planets in the modelled system. Significantly, the super frequencies from mean motion resonances, expected to induce TTV signals, match the significant peaks found by \citet{Ofir2018}. The super frequency of Kepler-82b/c corresponds to the main peak in the TTVs of Kepler-82b. Kepler-82c/f have a super frequency that explains the main peak of the TTVs in Kepler-82c. And finally, the super frequency of Kepler-82b/f matches a significant peak in the TTVs of Kepler-82c. Additionally the chopping frequency of Kepler-82c/f explains another peak of Kepler-82c TTVs. Two of the \citet{Ofir2018} frequencies with smaller confidence remain unexplained, these are the $(101.5\pm2.8)\cdot 10^{-4}\;\text{d}^{-1}$ frequency in planet b and the $(68.9\pm3.2)\cdot 10^{-4}\;\text{d}^{-1}$ frequency in planet c.

For comparison, we computed the same frequencies from the 6:2:1 resonance results. In this case the super frequency of Kepler-82b/c matches, as expected, with the main peak of the Kepler-82b frequencies, and the orbital frequency of the third non-transiting planet matches the $(58.9\pm3.2)\cdot10^{-4}\;\text{d}^{-1}$ peak. Besides these, no other matching frequencies were found, especially the main peak in the TTVs of Kepler-82c is not explained.

\subsection{Stellar parameters}
\label{subsec:mesa}
\begin{figure}
\resizebox{\hsize}{!}{\includegraphics{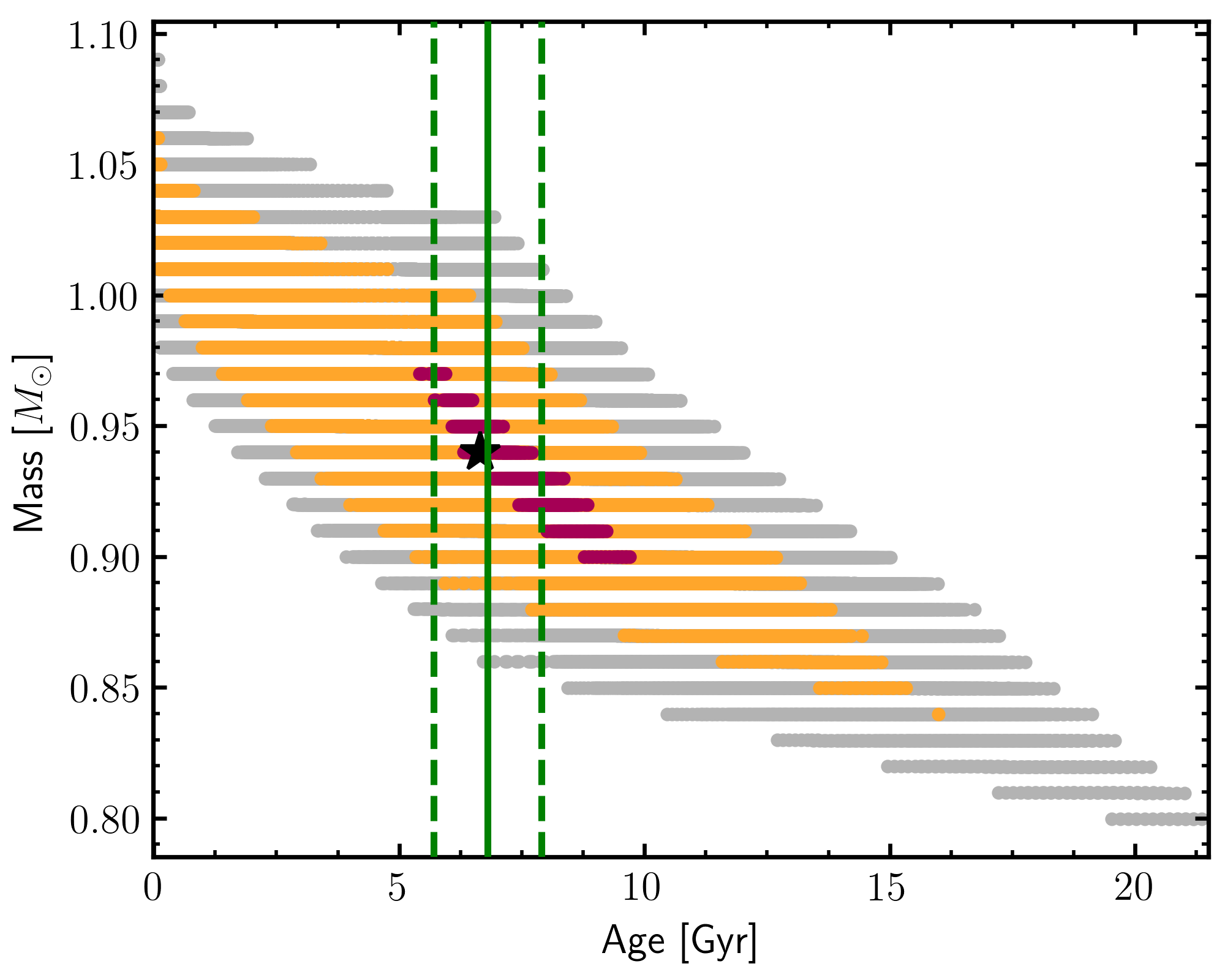}}
\caption{Mass-age diagram of Kepler-82 from MESA stellar evolution models (MIST). The black star and the red, orange, and grey dots correspond to the best matching value and the $1\sigma$, $2\sigma$, and $3\sigma$ areas derived from results on the density of the whole set photodynamical modelling and from the literature values of the effective temperature, the surface gravity, and the metallicity by \citet{Petigura2017}. The gyrochronologic age is indicated in green by a solid line and its 1-$\sigma$ range as dashed lines.}
\label{fig:MassAge2}
\end{figure}
\begin{figure}
\resizebox{\hsize}{!}{\includegraphics{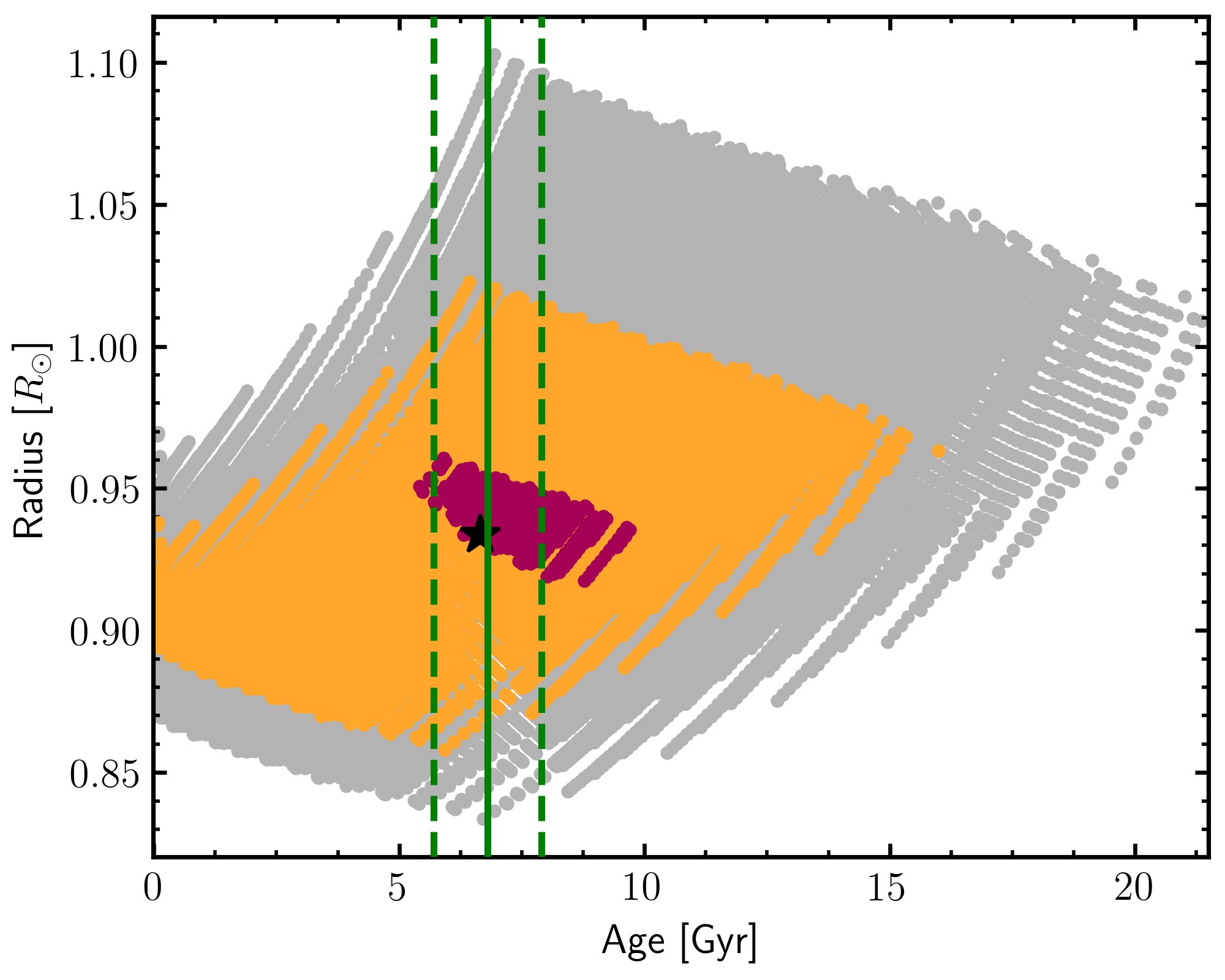}}
\caption{Radius-age diagram of Kepler-82 from MESA stellar evolution models (MIST). The black star and the red, orange, and grey dots correspond  to the best matching value and the $1\sigma$, $2\sigma$, and $3\sigma$ areas derived from results on the density of the whole data set photodynamical modelling and from the literature values of the effective temperature, the surface gravity, and the metallicity by \citet{Petigura2017}. The gyrochronologic age is indicated in green by a solid line and its 1-$\sigma$ range as dashed lines.}
\label{fig:RadiusAge2}
\end{figure}

Transit measurements provide the information about the stellar density \citep{AgolFabrycky2017}. In our photodynamical analysis we decided to model the stellar radius while fixing the stellar mass. With this parameterisation, the density is modelled as well. We derived the stellar radius to be $R_S=0.898_{-0.017}^{+0.038}\;R_\Sun$. The high asymmetry in the uncertainties is attributed to the symmetry of the inclination of Kepler-82c around $90^\degree$. Together with the stellar mass from \citet{Johnson2017} of $m_S=0.91$, this results in a stellar density of $\rho_S=1.77_{-0.21}^{+0.12} \text{ g cm}^{-3}$.

With this photodynamical-determined density and the measured stellar parameters \citep[from HIRES observations within the California-Kepler Survey]{Petigura2017} of the effective temperature $T_\text{eff}=5400.5\pm60\;\text{K}$, the surface gravity ${\log g =4.372\pm0.100}$, and metallicity $\text{Fe/H}=0.201\pm0.040$ we modelled the stellar radius, mass, and age with stellar evolution models. We extracted the corresponding values from MESA \citep{Paxton2011,Paxton2013,Paxton2015} evolutionary tracks interpolated by MIST \citep{Dotter2016,Choi2016}, rejecting values of the very early evolution below $0.1\;\text{Gyr}$. The results are visualised in Fig.~\ref{fig:MassAge2} as a mass-age diagram and in Fig.~\ref{fig:RadiusAge2} as a radius-age diagram with the best-matching value and the $1\sigma$, $2\sigma$, and $3\sigma$ areas as a black star and red, orange and grey dots respectively. For comparison, the gyrochronological stellar age derived below is plotted in green; it fits within the $1\sigma$ errorbars. The stellar parameters are derived to be $m_S=0.94_{-0.04}^{+0.03}\;M_\Sun$ for the mass, $R_S=0.934_{-0.016}^{+0.046}\;R_\Sun$ for the radius, and a stellar age of $\tau_\text{evol}=6.7_{-1.2}^{+3.0}\;$Gyr. 

\begin{figure}
\resizebox{\hsize}{!}{\includegraphics{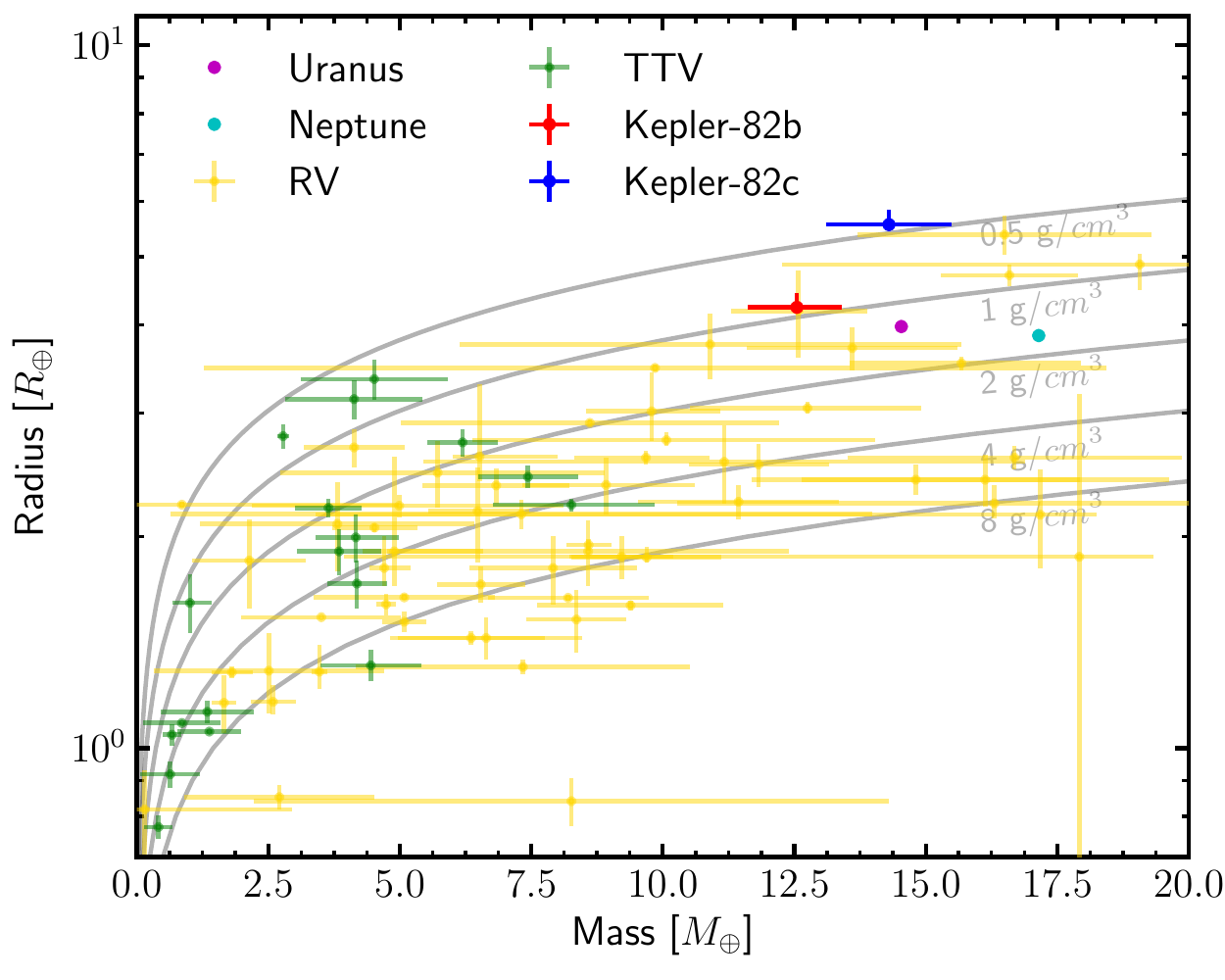}}
\caption{Mass-radius diagram for known planets with masses up to $20\;M_\Earth$. In yellow are the planets with mass measurements obtained by RVs and in green the planets with mass measurements obtained from TTVs. The data are given by The Extrasolar Planets Encyclopaedia. Our results are shown in red (Kepler-82b) and in blue (Kepler-82c). For comparison also the values of Uranus and Neptune are shown, the Neptune-like planet pair of our solar system.}
\label{fig:MassRadius}
\end{figure}

We corrected the photodynamically-determined parameters that depend on stellar mass and radius, namely planetary masses, semi-major axes, and radii, with these newly determined values. The corrected values are listed in column six of Table~\ref{table:results}. The planetary masses and radii of Kepler-82b/c are compared in Fig.~\ref{fig:MassRadius} with literature values of planets with masses up to $20\;M_\Earth$ from \texttt{The Extrasolar Planets Encyclopaedia\footnote{\url{http://exoplanet.eu/}}}.

For testing the results of the stellar evolution model analysis we applied the gyrochronologic age determination method to the Kepler-82 system. Therefore we determined its rotation period from the \Kepler long-cadence photometry excluding the transits of Kepler-82b/c \citep{Lomb,Scargle,LombScargle}. There are three small amplitude peaks in the periodogram; from these the highest-power peak corresponds to $34.7\pm0.8\;\text{days}$. Here, the period and error are determined as the mean and standard deviation from fitting a Gaussian to the peak. We made use of \citet{Barnes2007,Barnes2009} gyrochronologic estimation for determining the age of Kepler-82 based on its rotational period: 
\begin{equation}
\log(\tau_\text{Gyro}) = \frac{1}{n}[\log P - \log a - b \times \log (\text{B-V} - c)]\, ,
\end{equation}
with  $a = 0.770 \pm 0.014$, $b = 0.553 \pm 0.052$, $c = 0.472 \pm 0.027$, and $n = 0.519 \pm 0.007$. Assuming the spectral type G7 for Kepler-82 leads to B-V~$= 0.721$ \citep{Everett2012}. Following the \citet{Barnes2009} error estimation, we derive the gyrochronological age of Kepler-82 to be $6.8\pm1.1\;\text{Gyr}$. This value fits the age determined by stellar evolution models very well within the $1\sigma$ range. It is indicated with green in the mass-age and radius-age diagram (Fig.~\ref{fig:MassAge2} and \ref{fig:RadiusAge2}) with the mean as a solid line and the standard deviation in dashed lines.

From the recently published second \textit{Gaia} data release \citep{Gaia2016,Gaia2018} the effective temperature and the stellar radius were calculated to be $T_\text{eff}=5401\pm180\;\text{K}$ and $R_S=0.854_{-0.046}^{+0.043}\;R_\Sun$ by \citet{Berger2018}. While the effective temperature perfectly fits the HIRES value, the stellar radius is significantly smaller. It fits within the $1\sigma$ range of the value derived by the photodynamical analysis, and within the $2\sigma$ range the value derived with the stellar evolution models. The distance of Kepler-82 is determined to $905_{-22}^{+21}\;\text{pc}$ by \citet{Berger2018}.

The discrepancy between stellar parameters derived by GAIA and by the combination of the photodynamical analysis and spectroscopic parameters could be a hint of another star that contaminates the light of Kepler-82. Inspecting a small region around Kepler-82 revealed a star about two magnitudes fainter at 10~arcsec distance. This distance is large enough so that the \Kepler light curve should not be contaminated by this star. In the unlikely case of contamination, the radii of the planets would be underestimated by ten percent in maximum. That would make Kepler-82c to be an even more puffed-up exoplanet in the Neptune-like regime. The stellar radius and hence the density should not be affected by the light of a second star, as it is dependent upon the transit duration which is not changed. This agrees well with the fact that the photodynamical determined stellar radius matches the GAIA radius within its errorbars. It should be noted that the largest discrepancy is between the GAIA and the spectroscopic measurement, whereas the photodynamical one is in between. A further research of this deviance is beyond the scope of this paper.

\section{Conclusions}
\label{sec:conclusion}
In this work the first dynamical analysis of the Kepler-82 system was carried out, resulting in the discovery of a fifth planet. The signal of this planet is found in the TTVs of Kepler-82c. In addition to the sinusoidal behaviour due to the interaction with Kepler-82b being near the 2:1 resonance, the TTVs show the so called chopping signal manifesting in a jump every three consecutive transits. After optimising a 2-planet photodynamical model near the 2:1 resonance to the \Kepler long- and short-cadence data, we analysed the data with two different 3-planet system models. The systems differ in the ratio of the distance of the outermost fifth planet to Kepler-82c, either a 6:2:1 or a 3:2:1 near-resonant system were possible. The first evidence that the 3:2:1 resonance system model was the correct assumption was provided by the $\chi^2_\text{red}$, which is a little better than the one from analysing with the 6:2:1 resonance model. The 3:2:1 resonance model is also more favourable considering the mass of planet f is of the same order as planets b and c. This system model better fits into the 'peas in a pod' architecture of most systems found by \Kepler \citep{Weiss2018}. This is emphasised by the light curves collected in the framework of KOINet. The three new transit observations prefer the 3:2:1 resonance model and in addition a light curve including no transit measurement was taken during the time where a transit was predicted by the 6:2:1 resonance model. Additionally, the avoidance of inclinations that lead to transits by the third planet in near 3:2 resonance fits very well with the observations. Finally, with the periods of the planets in the 3:2:1 resonance system, except for two of them the frequencies in the TTVs of Kepler-82b/c detected by \citet{Ofir2018} can be explained by the super and the chopping frequencies. The most important point here is that \citet{Ofir2018} noticed a significant offset in the highest amplitude frequency of the TTVs of Kepler-82c from the near 2:1 mean motion resonance frequency. This peak is explained by the super frequency of the near 3:2 resonance of Kepler-82f and Kepler-82c. We conclude with announcing the detection of a fifth planet positioned in near 3:2 resonance to Kepler-82c. After the recent discovery of Kepler-411e \citep{Sun2019}, Kepler-82f is the second non-transiting planet detected via the TTVs of two other planets.

Determining the correct system architecture is important for modelling the right planet compositions. These highly depend on the assumed system architecture. The 2-planet model with significantly higher $\chi^2_\text{red}$ supposes a density ratio between planet b and c of about $\sim14$, whereas in the 3:2:1 resonance model especially the mass of Kepler-82b drops by about an order of magnitude resulting in a much more common (and reasonable) density ratio of $\sim2$.

\begin{acknowledgements}
We acknowledge funding from the German Research Foundation (DFG) through grant DR 281/30-1. This work made use of PyAstronomy. This research has made use of the NASA Exoplanet Archive, which is operated by the California Institute of Technology, under contract with the National Aeronautics and Space Administration under the Exoplanet Exploration Program. CvE acknowledges funding for the Stellar Astrophysics Centre, provided by The Danish National Research Foundation (Grant DNRF106), and support from the European Social Fund via the Lithuanian Science Council grant No. 09.3.3-LMT-K-712-01-0103. E.H, and I.R., acknowledge support from the Spanish Ministry of Economy and Competitiveness (MINECO) and the Fondo Europeo de Desarrollo Regional (FEDER) through grant ESP2016-80435-C2-1-R, as well as the support of the Generalitat de Catalunya/CERCA program. EA is supported by the United States National Science Foundation grant 1615315. Based on observations obtained with the Apache Point Observatory 3.5-meter telescope, which is owned and operated by the Astrophysical Research Consortium. We acknowledge support from the Research Council of Norway (grant 188910) to finance service observing at the NOT. SW acknowledges support for International Team 265 (``Magnetic Activity of M-type Dwarf Stars and the Influence on Habitable Extra-solar Planets'') funded by the International Space Science Institute (ISSI) in Bern, Switzerland, and by the Research Council of Norway through its Centres of Excellence scheme, project number 262622.

\end{acknowledgements}

\bibliography{references}
\bibliographystyle{aa.bst}

\begin{appendix}

\section{Additional plots and tables}

\longtab[1]{
\begin{longtable}{rllrllrllrll}
\caption{Ephemerides E and transit time predictions in BJD-2400000.0 from modelling \textit{Kepler} and KOINet data for the next 15 years. }\\
\label{table:ttv}
\\
\hline
\hline
E     & BJD              & & E    &     BJD          & & E    & BJD & & E    & BJD \\
\hline
\multicolumn{2}{l}{Kepler-82b:} & & &  & & &  & & & \\
\hline
$	133	$ & $	58491.2327	(	78	) $ & &$	188	$ & $	59945.6559	(	140	) $ & &$	243	$ & $	61400.0816	(	164	) $ & &$	298	$ & $	62854.5228	(	154	)$ \\
$	134	$ & $	58517.6757	(	80	) $ & &$	189	$ & $	59972.1003	(	141	) $ & &$	244	$ & $	61426.5237	(	164	) $ & &$	299	$ & $	62880.9705	(	152	)$ \\
$	135	$ & $	58544.1187	(	81	) $ & &$	190	$ & $	59998.5465	(	141	) $ & &$	245	$ & $	61452.9658	(	165	) $ & &$	300	$ & $	62907.4159	(	152	)$ \\
$	136	$ & $	58570.5609	(	83	) $ & &$	191	$ & $	60024.9908	(	141	) $ & &$	246	$ & $	61479.4095	(	165	) $ & &$	301	$ & $	62933.8627	(	151	)$ \\
$	137	$ & $	58597.0068	(	84	) $ & &$	192	$ & $	60051.4355	(	142	) $ & &$	247	$ & $	61505.8527	(	166	) $ & &$	302	$ & $	62960.3066	(	151	)$ \\
$	138	$ & $	58623.4495	(	86	) $ & &$	193	$ & $	60077.8793	(	143	) $ & &$	248	$ & $	61532.2962	(	166	) $ & &$	303	$ & $	62986.7544	(	150	)$ \\
$	139	$ & $	58649.8947	(	86	) $ & &$	194	$ & $	60104.3241	(	142	) $ & &$	249	$ & $	61558.7399	(	167	) $ & &$	304	$ & $	63013.1981	(	150	)$ \\
$	140	$ & $	58676.3391	(	88	) $ & &$	195	$ & $	60130.7667	(	143	) $ & &$	250	$ & $	61585.1852	(	167	) $ & &$	305	$ & $	63039.6433	(	149	)$ \\
$	141	$ & $	58702.7857	(	88	) $ & &$	196	$ & $	60157.2097	(	144	) $ & &$	251	$ & $	61611.6291	(	167	) $ & &$	306	$ & $	63066.0878	(	149	)$ \\
$	142	$ & $	58729.2287	(	90	) $ & &$	197	$ & $	60183.6534	(	145	) $ & &$	252	$ & $	61638.0751	(	167	) $ & &$	307	$ & $	63092.5323	(	148	)$ \\
$	143	$ & $	58755.6765	(	90	) $ & &$	198	$ & $	60210.0953	(	145	) $ & &$	253	$ & $	61664.5197	(	168	) $ & &$	308	$ & $	63118.9747	(	148	)$ \\
$	144	$ & $	58782.1205	(	92	) $ & &$	199	$ & $	60236.5371	(	147	) $ & &$	254	$ & $	61690.9672	(	167	) $ & &$	309	$ & $	63145.4188	(	148	)$ \\
$	145	$ & $	58808.5676	(	91	) $ & &$	200	$ & $	60262.9801	(	148	) $ & &$	255	$ & $	61717.4115	(	168	) $ & &$	310	$ & $	63171.8614	(	149	)$ \\
$	146	$ & $	58835.0118	(	93	) $ & &$	201	$ & $	60289.4217	(	148	) $ & &$	256	$ & $	61743.8595	(	167	) $ & &$	311	$ & $	63198.3035	(	149	)$ \\
$	147	$ & $	58861.4591	(	93	) $ & &$	202	$ & $	60315.8630	(	150	) $ & &$	257	$ & $	61770.3053	(	167	) $ & &$	312	$ & $	63224.7456	(	149	)$ \\
$	148	$ & $	58887.9031	(	94	) $ & &$	203	$ & $	60342.3056	(	151	) $ & &$	258	$ & $	61796.7530	(	166	) $ & &$	313	$ & $	63251.1875	(	150	)$ \\
$	149	$ & $	58914.3495	(	94	) $ & &$	204	$ & $	60368.7475	(	152	) $ & &$	259	$ & $	61823.1979	(	166	) $ & &$	314	$ & $	63277.6289	(	150	)$ \\
$	150	$ & $	58940.7934	(	96	) $ & &$	205	$ & $	60395.1890	(	153	) $ & &$	260	$ & $	61849.6473	(	164	) $ & &$	315	$ & $	63304.0698	(	151	)$ \\
$	151	$ & $	58967.2398	(	95	) $ & &$	206	$ & $	60421.6317	(	155	) $ & &$	261	$ & $	61876.0919	(	165	) $ & &$	316	$ & $	63330.5110	(	152	)$ \\
$	152	$ & $	58993.6827	(	97	) $ & &$	207	$ & $	60448.0743	(	156	) $ & &$	262	$ & $	61902.5393	(	163	) $ & &$	317	$ & $	63356.9527	(	152	)$ \\
$	153	$ & $	59020.1276	(	97	) $ & &$	208	$ & $	60474.5168	(	157	) $ & &$	263	$ & $	61928.9854	(	163	) $ & &$	318	$ & $	63383.3930	(	154	)$ \\
$	154	$ & $	59046.5716	(	98	) $ & &$	209	$ & $	60500.9603	(	158	) $ & &$	264	$ & $	61955.4328	(	161	) $ & &$	319	$ & $	63409.8340	(	155	)$ \\
$	155	$ & $	59073.0148	(	99	) $ & &$	210	$ & $	60527.4037	(	159	) $ & &$	265	$ & $	61981.8771	(	161	) $ & &$	320	$ & $	63436.2758	(	155	)$ \\
$	156	$ & $	59099.4571	(	100	) $ & &$	211	$ & $	60553.8485	(	160	) $ & &$	266	$ & $	62008.3245	(	160	) $ & &$	321	$ & $	63462.7167	(	156	)$ \\
$	157	$ & $	59125.9010	(	101	) $ & &$	212	$ & $	60580.2918	(	162	) $ & &$	267	$ & $	62034.7689	(	159	) $ & &$	322	$ & $	63489.1580	(	158	)$ \\
$	158	$ & $	59152.3426	(	102	) $ & &$	213	$ & $	60606.7375	(	162	) $ & &$	268	$ & $	62061.2140	(	158	) $ & &$	323	$ & $	63515.6009	(	159	)$ \\
$	159	$ & $	59178.7840	(	104	) $ & &$	214	$ & $	60633.1824	(	163	) $ & &$	269	$ & $	62087.6584	(	158	) $ & &$	324	$ & $	63542.0429	(	160	)$ \\
$	160	$ & $	59205.2266	(	105	) $ & &$	215	$ & $	60659.6285	(	163	) $ & &$	270	$ & $	62114.1031	(	157	) $ & &$	325	$ & $	63568.4846	(	161	)$ \\
$	161	$ & $	59231.6677	(	107	) $ & &$	216	$ & $	60686.0730	(	165	) $ & &$	271	$ & $	62140.5463	(	156	) $ & &$	326	$ & $	63594.9293	(	161	)$ \\
$	162	$ & $	59258.1085	(	108	) $ & &$	217	$ & $	60712.5216	(	164	) $ & &$	272	$ & $	62166.9898	(	156	) $ & &$	327	$ & $	63621.3717	(	162	)$ \\
$	163	$ & $	59284.5502	(	110	) $ & &$	218	$ & $	60738.9659	(	165	) $ & &$	273	$ & $	62193.4328	(	155	) $ & &$	328	$ & $	63647.8158	(	163	)$ \\
$	164	$ & $	59310.9914	(	112	) $ & &$	219	$ & $	60765.4137	(	165	) $ & &$	274	$ & $	62219.8757	(	154	) $ & &$	329	$ & $	63674.2594	(	164	)$ \\
$	165	$ & $	59337.4321	(	114	) $ & &$	220	$ & $	60791.8599	(	166	) $ & &$	275	$ & $	62246.3176	(	154	) $ & &$	330	$ & $	63700.7052	(	164	)$ \\
$	166	$ & $	59363.8735	(	116	) $ & &$	221	$ & $	60818.3081	(	164	) $ & &$	276	$ & $	62272.7598	(	155	) $ & &$	331	$ & $	63727.1480	(	165	)$ \\
$	167	$ & $	59390.3151	(	117	) $ & &$	222	$ & $	60844.7527	(	166	) $ & &$	277	$ & $	62299.2023	(	154	) $ & &$	332	$ & $	63753.5951	(	165	)$ \\
$	168	$ & $	59416.7567	(	119	) $ & &$	223	$ & $	60871.2021	(	165	) $ & &$	278	$ & $	62325.6436	(	154	) $ & &$	333	$ & $	63780.0389	(	166	)$ \\
$	169	$ & $	59443.1983	(	121	) $ & &$	224	$ & $	60897.6472	(	165	) $ & &$	279	$ & $	62352.0850	(	154	) $ & &$	334	$ & $	63806.4861	(	165	)$ \\
$	170	$ & $	59469.6408	(	123	) $ & &$	225	$ & $	60924.0946	(	164	) $ & &$	280	$ & $	62378.5279	(	154	) $ & &$	335	$ & $	63832.9296	(	167	)$ \\
$	171	$ & $	59496.0836	(	125	) $ & &$	226	$ & $	60950.5402	(	165	) $ & &$	281	$ & $	62404.9690	(	154	) $ & &$	336	$ & $	63859.3772	(	166	)$ \\
$	172	$ & $	59522.5265	(	126	) $ & &$	227	$ & $	60976.9875	(	163	) $ & &$	282	$ & $	62431.4105	(	155	) $ & &$	337	$ & $	63885.8215	(	167	)$ \\
$	173	$ & $	59548.9693	(	128	) $ & &$	228	$ & $	61003.4318	(	163	) $ & &$	283	$ & $	62457.8536	(	155	) $ & &$	338	$ & $	63912.2682	(	166	)$ \\
$	174	$ & $	59575.4148	(	129	) $ & &$	229	$ & $	61029.8781	(	163	) $ & &$	284	$ & $	62484.2956	(	155	) $ & &$	339	$ & $	63938.7120	(	167	)$ \\
$	175	$ & $	59601.8574	(	131	) $ & &$	230	$ & $	61056.3226	(	163	) $ & &$	285	$ & $	62510.7379	(	155	) $ & &$	340	$ & $	63965.1594	(	167	)$ \\
$	176	$ & $	59628.3030	(	132	) $ & &$	231	$ & $	61082.7678	(	162	) $ & &$	286	$ & $	62537.1817	(	156	) $ & &$	341	$ & $	63991.6023	(	168	)$ \\
$	177	$ & $	59654.7478	(	134	) $ & &$	232	$ & $	61109.2114	(	162	) $ & &$	287	$ & $	62563.6254	(	156	) $ & &$	342	$ & $	64018.0479	(	167	)$ \\
$	178	$ & $	59681.1942	(	134	) $ & &$	233	$ & $	61135.6556	(	162	) $ & &$	288	$ & $	62590.0680	(	156	) $ & &$	343	$ & $	64044.4922	(	168	)$ \\
$	179	$ & $	59707.6379	(	136	) $ & &$	234	$ & $	61162.0995	(	161	) $ & &$	289	$ & $	62616.5136	(	156	) $ & &$	344	$ & $	64070.9364	(	168	)$ \\
$	180	$ & $	59734.0866	(	136	) $ & &$	235	$ & $	61188.5422	(	162	) $ & &$	290	$ & $	62642.9573	(	156	) $ & &$	345	$ & $	64097.3789	(	169	)$ \\
$	181	$ & $	59760.5308	(	137	) $ & &$	236	$ & $	61214.9851	(	162	) $ & &$	291	$ & $	62669.4031	(	155	) $ & &$	346	$ & $	64123.8236	(	170	)$ \\
$	182	$ & $	59786.9783	(	137	) $ & &$	237	$ & $	61241.4285	(	162	) $ & &$	292	$ & $	62695.8471	(	156	) $ & &$	347	$ & $	64150.2657	(	171	)$ \\
$	183	$ & $	59813.4238	(	139	) $ & &$	238	$ & $	61267.8702	(	162	) $ & &$	293	$ & $	62722.2942	(	155	) $ & &$	348	$ & $	64176.7075	(	172	)$ \\
$	184	$ & $	59839.8718	(	138	) $ & &$	239	$ & $	61294.3120	(	162	) $ & &$	294	$ & $	62748.7387	(	155	) $ & &$	349	$ & $	64203.1501	(	173	)$ \\
$	185	$ & $	59866.3161	(	140	) $ & &$	240	$ & $	61320.7550	(	162	) $ & &$	295	$ & $	62775.1860	(	155	) $ & &$	350	$ & $	64229.5917	(	174	)$ \\
$	186	$ & $	59892.7641	(	139	) $ & &$	241	$ & $	61347.1967	(	163	) $ & &$	296	$ & $	62801.6304	(	155	) $ & &$	351	$ & $	64256.0326	(	175	)$ \\
$	187	$ & $	59919.2089	(	141	) $ & &$	242	$ & $	61373.6382	(	163	) $ & &$	297	$ & $	62828.0789	(	153	) $ & &$	352	$ & $	64282.4739	(	177	)$ \\
\hline
\caption*{{\bf Notes}. The median and standard deviation solution of 1000 randomly chosen good models. Reference times for ephemeris E=1: ${T_b=54974.2409(12)}$ and $T_c=54955.5862(22)$}\\
\\
\caption{continued.}\\
\hline
\hline
E     & BJD              & & E    &     BJD          & & E    & BJD & & E    & BJD \\
\hline       
\multicolumn{2}{l}{Kepler-82c:} & & &  & & &  & & &\\
\hline
$	69	$ & $	58511.9213	(	51	) $ & &$	97	$ & $	59955.0537	(	94	) $ & &$	125	$ & $	61398.0596	(	170	) $ & &$	153	$ & $	62841.1069	(	181	)$ \\
$	70	$ & $	58563.4587	(	50	) $ & &$	98	$ & $	60006.5884	(	106	) $ & &$	126	$ & $	61449.5997	(	168	) $ & &$	154	$ & $	62892.6426	(	169	)$ \\
$	71	$ & $	58615.0271	(	63	) $ & &$	99	$ & $	60058.1120	(	99	) $ & &$	127	$ & $	61501.1194	(	171	) $ & &$	155	$ & $	62944.1653	(	168	)$ \\
$	72	$ & $	58666.5477	(	58	) $ & &$	100	$ & $	60109.6316	(	109	) $ & &$	128	$ & $	61552.6564	(	211	) $ & &$	156	$ & $	62995.7474	(	207	)$ \\
$	73	$ & $	58718.0834	(	51	) $ & &$	101	$ & $	60161.1717	(	114	) $ & &$	129	$ & $	61604.1925	(	205	) $ & &$	157	$ & $	63047.2834	(	187	)$ \\
$	74	$ & $	58769.6185	(	58	) $ & &$	102	$ & $	60212.6931	(	118	) $ & &$	130	$ & $	61655.7116	(	210	) $ & &$	158	$ & $	63098.8153	(	183	)$ \\
$	75	$ & $	58821.1380	(	52	) $ & &$	103	$ & $	60264.2227	(	140	) $ & &$	131	$ & $	61707.2758	(	266	) $ & &$	159	$ & $	63150.4148	(	180	)$ \\
$	76	$ & $	58872.6773	(	49	) $ & &$	104	$ & $	60315.7634	(	137	) $ & &$	132	$ & $	61758.8083	(	257	) $ & &$	160	$ & $	63201.9487	(	157	)$ \\
$	77	$ & $	58924.1995	(	49	) $ & &$	105	$ & $	60367.2860	(	141	) $ & &$	133	$ & $	61810.3307	(	263	) $ & &$	161	$ & $	63253.4866	(	151	)$ \\
$	78	$ & $	58975.7233	(	48	) $ & &$	106	$ & $	60418.8402	(	184	) $ & &$	134	$ & $	61861.9183	(	295	) $ & &$	162	$ & $	63305.0819	(	114	)$ \\
$	79	$ & $	59027.2653	(	47	) $ & &$	107	$ & $	60470.3785	(	177	) $ & &$	135	$ & $	61913.4470	(	279	) $ & &$	163	$ & $	63356.6105	(	113	)$ \\
$	80	$ & $	59078.7886	(	49	) $ & &$	108	$ & $	60521.9032	(	183	) $ & &$	136	$ & $	61964.9745	(	281	) $ & &$	164	$ & $	63408.1497	(	119	)$ \\
$	81	$ & $	59130.3301	(	62	) $ & &$	109	$ & $	60573.4843	(	225	) $ & &$	137	$ & $	62016.5681	(	261	) $ & &$	165	$ & $	63459.7170	(	151	)$ \\
$	82	$ & $	59181.8732	(	58	) $ & &$	110	$ & $	60625.0176	(	219	) $ & &$	138	$ & $	62068.0914	(	244	) $ & &$	166	$ & $	63511.2413	(	161	)$ \\
$	83	$ & $	59233.4008	(	59	) $ & &$	111	$ & $	60676.5439	(	228	) $ & &$	139	$ & $	62119.6225	(	243	) $ & &$	167	$ & $	63562.7815	(	162	)$ \\
$	84	$ & $	59284.9733	(	87	) $ & &$	112	$ & $	60728.1367	(	240	) $ & &$	140	$ & $	62171.2001	(	178	) $ & &$	168	$ & $	63614.3177	(	210	)$ \\
$	85	$ & $	59336.5156	(	80	) $ & &$	113	$ & $	60779.6613	(	234	) $ & &$	141	$ & $	62222.7205	(	171	) $ & &$	169	$ & $	63665.8397	(	210	)$ \\
$	86	$ & $	59388.0482	(	83	) $ & &$	114	$ & $	60831.1886	(	243	) $ & &$	142	$ & $	62274.2547	(	175	) $ & &$	170	$ & $	63717.3808	(	207	)$ \\
$	87	$ & $	59439.6448	(	102	) $ & &$	115	$ & $	60882.7736	(	213	) $ & &$	143	$ & $	62325.8041	(	127	) $ & &$	171	$ & $	63768.9029	(	219	)$ \\
$	88	$ & $	59491.1819	(	98	) $ & &$	116	$ & $	60934.2908	(	211	) $ & &$	144	$ & $	62377.3249	(	129	) $ & &$	172	$ & $	63820.4267	(	202	)$ \\
$	89	$ & $	59542.7160	(	107	) $ & &$	117	$ & $	60985.8186	(	220	) $ & &$	145	$ & $	62428.8641	(	135	) $ & &$	173	$ & $	63871.9669	(	202	)$ \\
$	90	$ & $	59594.3141	(	102	) $ & &$	118	$ & $	61037.3776	(	167	) $ & &$	146	$ & $	62480.3915	(	129	) $ & &$	174	$ & $	63923.4873	(	200	)$ \\
$	91	$ & $	59645.8400	(	104	) $ & &$	119	$ & $	61088.8929	(	171	) $ & &$	147	$ & $	62531.9134	(	133	) $ & &$	175	$ & $	63975.0305	(	161	)$ \\
$	92	$ & $	59697.3710	(	116	) $ & &$	120	$ & $	61140.4255	(	182	) $ & &$	148	$ & $	62583.4535	(	135	) $ & &$	176	$ & $	64026.5707	(	159	)$ \\
$	93	$ & $	59748.9480	(	90	) $ & &$	121	$ & $	61191.9578	(	149	) $ & &$	149	$ & $	62634.9733	(	137	) $ & &$	177	$ & $	64078.0964	(	152	)$ \\
$	94	$ & $	59800.4643	(	96	) $ & &$	122	$ & $	61243.4763	(	155	) $ & &$	150	$ & $	62686.5010	(	148	) $ & &$	178	$ & $	64129.6718	(	140	)$ \\
$	95	$ & $	59851.9942	(	111	) $ & &$	123	$ & $	61295.0151	(	161	) $ & &$	151	$ & $	62738.0382	(	144	) $ & &$	179	$ & $	64181.2144	(	132	)$ \\
$	96	$ & $	59903.5382	(	85	) $ & &$	124	$ & $	61346.5365	(	157	) $ & &$	152	$ & $	62789.5566	(	144	) $ & &$	180	$ & $	64232.7492	(	127	)$ \\
\hline
\end{longtable}
}
\longtab[2]{
\begin{longtable}{l c c c c c}
\caption{Stellar and planetary parameters from the photodynamical modelling of the 2:1 resonance solution on \Kepler long- and short-cadence data, the 6:2:1 resonance solution on \Kepler data, the 3:2:1 resonance solution on \Kepler data, the 3:2:1 resonance solution on \Kepler data and the three KOINet transit light curves, and some corrections from investigating stellar evolution models in Sect.~\ref{subsec:mesa}.} \\
\label{table:results}
\\
\hline
\hline															
Parameter & Kepler data &   Kepler data &   Kepler data & Kepler \& & MESA  \\
&   & & & KOINet data & \\
&   2:1   &	6:2:1   &   3:2:1    &   3:2:1 &    3:2:1	\\
\hline															
\noalign{\smallskip}																															
\multicolumn{5}{l}{\bf Kepler-82b}																									&						\\
$m_b/m_S$	& $	0.000682	_{-	0.000022	}^{+	0.000021	}$ & $	0.0000563	_{-	0.0000076	}^{+	0.0000072	}$ & $	0.0000374	_{-	0.0000055	}^{+	0.0000069	}$ & $	0.0000401	_{-	0.0000025	}^{+	0.0000028	}$ &						\\
$m_b^* \;[M_\Earth]$	& $	207			\pm	10	$ & $	17.1	_{-	2.4	}^{+	2.3	}$ & $	11.3	_{-	1.7	}^{+	2.1	}$ & $	12.15	_{-	0.87	}^{+	0.96	}$ & $	12.5	_{-	1.0	}^{+	0.9	}$ \\
$a_{b,\text{corr}}$	& $	0.9999097	_{-	0.0000066	}^{+	0.0000067	}$ & $	0.999927			\pm	0.000011	$ & $	0.9999605	_{-	0.0000058	}^{+	0.0000057	}$ & $	0.9999606	_{-	0.0000041	}^{+	0.0000042	}$ &						\\
$a_b^* \;[\text{AU}]$	& $	0.1684			\pm	0.0021	$ & $	0.1683			\pm	0.0020	$ & $	0.1683			\pm	0.0020	$ & $	0.1683			\pm	0.0020	$ & $	0.1702	_{-	0.0024	}^{+	0.0018	}$ \\
$P_b^* \;[\text{d}]$	& $	26.44			\pm	0.48	$ & $	26.44			\pm	0.48	$ & $	26.44			\pm	0.48	$ & $	26.44			\pm	0.48	$ & $	26.44	_{-	0.56	}^{+	0.42	}$ \\
$e_b$	& $	0.01737	_{-	0.00067	}^{+	0.00068	}$ & $	0.0024	_{-	0.0013	}^{+	0.0011	}$ & $	0.0030	_{-	0.0019	}^{+	0.0024	}$ & $	0.0033	_{-	0.0017	}^{+	0.0019	}$ &						\\
$i_b\;[^\degree]$	& $	88.49	_{-	0.16	}^{+	0.18	}$ & $	89.109	_{-	0.066	}^{+	0.032	}$ & $	89.031	_{-	0.096	}^{+	0.068	}$ & $	89.052	_{-	0.096	}^{+	0.049	}$ &						\\
$\Omega_b\;[^\degree]$	& 	\multicolumn{4}{c}{$0$ (fixed)}																							&						\\
$\omega_b\;[^\degree]$	& $	254.6			\pm	1.3	$ & $	214	_{-	43	}^{+	37	}$ & $	231	_{-	46	}^{+	29	}$ & $	236	_{-	26	}^{+	23	}$ &						\\
$M_{b,\text{corr}}\;[^\degree]$	& $	0.079			\pm	0.018	$ & $	-0.094	_{-	0.021	}^{+	0.023	}$ & $	-0.026	_{-	0.021	}^{+	0.019	}$ & $	-0.025	_{-	0.020	}^{+	0.019	}$ &						\\
$M_b^* \;[^\degree]$	& $	354.6			\pm	1.3	$ & $	35	_{-	37	}^{+	43	}$ & $	18	_{-	29	}^{+	46	}$ & $	13	_{-	23	}^{+	26	}$ &						\\
$R_b/R_S $	& $	0.04428	_{-	0.00135	}^{+	0.00127	}$ & $	0.04125	_{-	0.00039	}^{+	0.00040	}$ & $	0.04166	_{-	0.00050	}^{+	0.00053	}$ & $	0.04159	_{-	0.00045	}^{+	0.00049	}$ &						\\
$R_b^* \;[R_\Earth]$	& $	5.74	_{-	0.57	}^{+	0.54	}$ & $	3.96	_{-	0.06	}^{+	0.15	}$ & $	4.13	_{-	0.16	}^{+	0.24	}$ & $	4.07	_{-	0.10	}^{+	0.24	}$ & $	4.24	_{-	0.09	}^{+	0.22	}$ \\
$\rho_b^* [\text{g\;cm}^{-3}]$	& $	6.0	_{-	1.4	}^{+	2.2	}$ & $	1.49	_{-	0.21	}^{+	0.19	}$ & $	0.88	_{-	0.17	}^{+	0.19	}$ & $	0.98	_{-	0.16	}^{+	0.11	}$ &						\\
\noalign{\smallskip}																															
\multicolumn{5}{l}{\bf Kepler-82c}																									&						\\
$m_c/m_b$	& $	0.156			\pm	0.012	$ & $	1.06			\pm	0.14	$ & $	1.23	_{-	0.20	}^{+	0.21	}$ & $	1.14	_{-	0.13	}^{+	0.14	}$ &						\\
$m_c^* \;[M_\Earth]$	& $	32.2			\pm	2.2	$ & $	18.0	_{-	2.0	}^{+	1.8	}$ & $	13.9	_{-	1.6	}^{+	1.4	}$ & $	13.9	_{-	1.2	}^{+	1.3	}$ & $	14.3			\pm	1.3	$ \\
$a_{c,\text{corr}}$	& $	1.003856	_{-	0.000099	}^{+	0.000095	}$ & $	0.999860	_{-	0.000044	}^{+	0.000041	}$ & $	0.999934	_{-	0.000029	}^{+	0.000030	}$ & $	0.999947	_{-	0.000019	}^{+	0.000018	}$ &						\\
$a_c^* \;[\text{AU}]$	& $	0.2637			\pm	0.0032	$ & $	0.2626			\pm	0.0032	$ & $	0.2626			\pm	0.0032	$ & $	0.2626			\pm	0.0032	$ & $	0.2655	_{-	0.0038	}^{+	0.0028	}$ \\
$P_c^* \;[\text{d}]$	& $	51.84			\pm	0.95	$ & $	51.53			\pm	0.94	$ & $	51.53			\pm	0.94	$ & $	51.54			\pm	0.94	$ & $	51.5	_{-	1.1	}^{+	0.8	}$ \\
$e_c$	& $	0.0685	_{-	0.0014	}^{+	0.0015	}$ & $	0.0103	_{-	0.0034	}^{+	0.0030	}$ & $	0.0072	_{-	0.0017	}^{+	0.0019	}$ & $	0.0070	_{-	0.0018	}^{+	0.0016	}$ &						\\
$i_c\;[^\degree]$ config. I	& $	89.19	_{-	0.10	}^{+	0.13	}$ & $	89.96			\pm	0.18	$ & $	90.18	_{-	0.18	}^{+	0.17	}$ & $	90.15	_{-	0.22	}^{+	0.18	}$ &						\\
$i_c\;[^\degree]$ config. II	& $	90.76	_{-	0.13	}^{+	0.12	}$ &	-					& $	89.73	_{-	0.12	}^{+	0.14	}$ & $	89.78	_{-	0.15	}^{+	0.17	}$ &						\\
$\Omega_c\;[^\degree]$ config. I	& $	0.01	_{-	0.10	}^{+	0.09	}$ & $	1.6	_{-	1.4	}^{+	1.3	}$ & $	1.8	_{-	2.1	}^{+	2.0	}$ & $	1.6			\pm	2.1	$ &						\\
$\Omega_c\;[^\degree]$ config. II	& $	0.29			\pm	0.10	$ & 	-					&	-					&		-				&						\\
$\omega_c\;[^\degree]$	& $	271.50	_{-	0.67	}^{+	0.69	}$ & $	268	_{-	16	}^{+	13	}$ & $	161	_{-	24	}^{+	12	}$ & $	162	_{-	20	}^{+	12	}$ &						\\
$M_{c,\text{corr}}\;[^\degree]$	& $	-0.952	_{-	0.044	}^{+	0.042	}$ & $	-0.554	_{-	0.019	}^{+	0.018	}$ & $	-0.509	_{-	0.019	}^{+	0.020	}$ & $	-0.507			\pm	0.020	$ &						\\
$M_c^* \;[^\degree]$	& $	19.99	_{-	0.78	}^{+	0.75	}$ & $	25	_{-	13	}^{+	16	}$ & $	132	_{-	12	}^{+	23	}$ & $	131	_{-	12	}^{+	20	}$ &						\\
$R_c/R_S $	& $	0.0578	_{-	0.0019	}^{+	0.0017	}$ & $	0.05423	_{-	0.00047	}^{+	0.00048	}$ & $	0.05461	_{-	0.00053	}^{+	0.00073	}$ & $	0.05453	_{-	0.00053	}^{+	0.00068	}$ &						\\
$R_c^* \;[R_\Earth]$	& $	7.49	_{-	0.77	}^{+	0.72	}$ & $	5.19	_{-	0.06	}^{+	0.20	}$ & $	5.41	_{-	0.20	}^{+	0.33	}$ & $	5.34	_{-	0.13	}^{+	0.32	}$ & $	5.56	_{-	0.11	}^{+	0.28	}$ \\
$\rho_c^* [\text{g\;cm}^{-3}]$	& $	0.42	_{-	0.11	}^{+	0.16	}$ & $	0.693	_{-	0.083	}^{+	0.077	}$ & $	0.480	_{-	0.088	}^{+	0.085	}$ & $	0.494	_{-	0.083	}^{+	0.070	}$ &						\\
\noalign{\smallskip}																															
\multicolumn{5}{l}{\bf Kepler-82f}																									&						\\
$m_f/m_c$	&	-					& $	23.0	_{-	2.1	}^{+	2.8	}$ & $	1.50	_{-	0.16	}^{+	0.17	}$ & $	1.50	_{-	0.13	}^{+	0.16	}$ &						\\
$m_f^* \;[M_\Earth]$	&	-					& $	415			\pm	23	$ & $	20.9			\pm	1.0	$ & $	20.9			\pm	1.0	$ & $	21.6	_{-	1.2	}^{+	1.0	}$ \\
$P_f/P_c$	&	-					& $	3.1245	_{-	0.0027	}^{+	0.0023	}$ & $	1.46969	_{-	0.00041	}^{+	0.00044	}$ & $	1.46940	_{-	0.00022	}^{+	0.00023	}$ &						\\
$P_f^*$ [d]	&	-					& $	161.03	_{-	0.14	}^{+	0.12	}$ & $	75.747	_{-	0.021	}^{+	0.023	}$ & $	75.732			\pm	0.012	$ &						\\
$a_f^* \;[\text{AU}]$	&	-					& $	0.5616			\pm	0.0068	$ & $	0.3395			\pm	0.0041	$ & $	0.3395			\pm	0.0041	$ & $	0.3432	_{-	0.0049	}^{+	0.0037	}$ \\
$e_f$	&	-					& $	0.0912	_{-	0.0047	}^{+	0.0066	}$ & $	0.0016	_{-	0.0013	}^{+	0.0022	}$ & $	0.0014	_{-	0.0010	}^{+	0.0018	}$ &						\\
$i_f\;[^\degree]$ config. I	&	-					& $	90.6	_{-	2.9	}^{+	3.7	}$ & $	86.21	_{-	0.68	}^{+	0.78	}$ & $	86.30			\pm	0.56	$ &						\\
$i_f\;[^\degree]$ config. II	&	-					& 	-					& $	93.63	_{-	0.67	}^{+	0.61	}$ & $	93.62	_{-	0.72	}^{+	0.56	}$ &						\\
$\Omega_f\;[^\degree]$ 	&	-					& $	1.7	_{-	1.7	}^{+	1.2	}$ & $	1.9			\pm	2.2	$ & $	1.6	_{-	2.1	}^{+	2.2	}$ &						\\
$\omega_f\;[^\degree]$	&	-					& $	279.5	_{-	4.7	}^{+	4.3	}$ & $	77	_{-	62	}^{+	67	}$ & $	62	_{-	47	}^{+	70	}$ &						\\
$M_f \;[^\degree]$	&	-					& $	92.5	_{-	4.4	}^{+	4.5	}$ & $	111	_{-	67	}^{+	61	}$ & $	125	_{-	70	}^{+	47	}$ &						\\
\hline
\caption*{{\bf Notes}. Listed are the median values and $68.26$\% confidence interval from the MCMC posterior distribution. The osculating orbital elements are given at a reference time, BJD = 2454933.0. $^{(*)}$Derived, not fitted parameters.}\\
\\
\caption{continued.}\\
\hline
\hline															
Parameter & Kepler data &   Kepler data &   Kepler data & Kepler \& & MESA  \\
&   & & & KOINet data & \\
&   2:1   &	6:2:1   &   3:2:1    &   3:2:1 &    3:2:1	\\
\hline															
\noalign{\smallskip}
\multicolumn{5}{l}{\bf Kepler-82}																									&						\\
$m_S [M_\Sun]$	&	\multicolumn{4}{c}{$0.91\pm0.03$ \citep[fixed,][]{Johnson2017}} 																							& $	0.94	_{-	0.04	}^{+	0.03	}$ \\
$R_S\;[R_\sun]$	& $	1.186	_{-	0.085	}^{+	0.082	}$ & $	0.880	_{-	0.010	}^{+	0.025	}$ & $	0.907	_{-	0.026	}^{+	0.042	}$ & $	0.898	_{-	0.018	}^{+	0.042	}$ & $	0.934	_{-	0.016	}^{+	0.046	}$ \\
$\rho_S^* [\text{g\,cm}^{-3}]$	& $	0.77	_{-	0.14	}^{+	0.19	}$ & $	1.89	_{-	0.15	}^{+	0.06	}$ & $	1.72	_{-	0.22	}^{+	0.16	}$ & $	1.77	_{-	0.23	}^{+	0.11	}$ &						\\
$c_{1,\text{\it Kepler}}$	& $	0.31	_{-	0.19	}^{+	0.23	}$ & $	0.498	_{-	0.074	}^{+	0.064	}$ & $	0.515	_{-	0.062	}^{+	0.048	}$ & $	0.522	_{-	0.075	}^{+	0.054	}$ &						\\
$c_{2,\text{\it Kepler}}$	& $	0.66	_{-	0.36	}^{+	0.26	}$ & $	0.19	_{-	0.11	}^{+	0.13	}$ & $	0.134	_{-	0.073	}^{+	0.090	}$ & $	0.12	_{-	0.09	}^{+	0.14	}$ &						\\
\hline
\caption*{{\bf Notes}. Listed are the median values and $68.26$\% confidence interval from the MCMC posterior distribution. The osculating orbital elements are given at a reference time, BJD = 2454933.0. $^{(*)}$Derived, not fitted parameters.}\\
\end{longtable}
}

\begin{figure*}
\sidecaption
  \includegraphics[width=12cm]{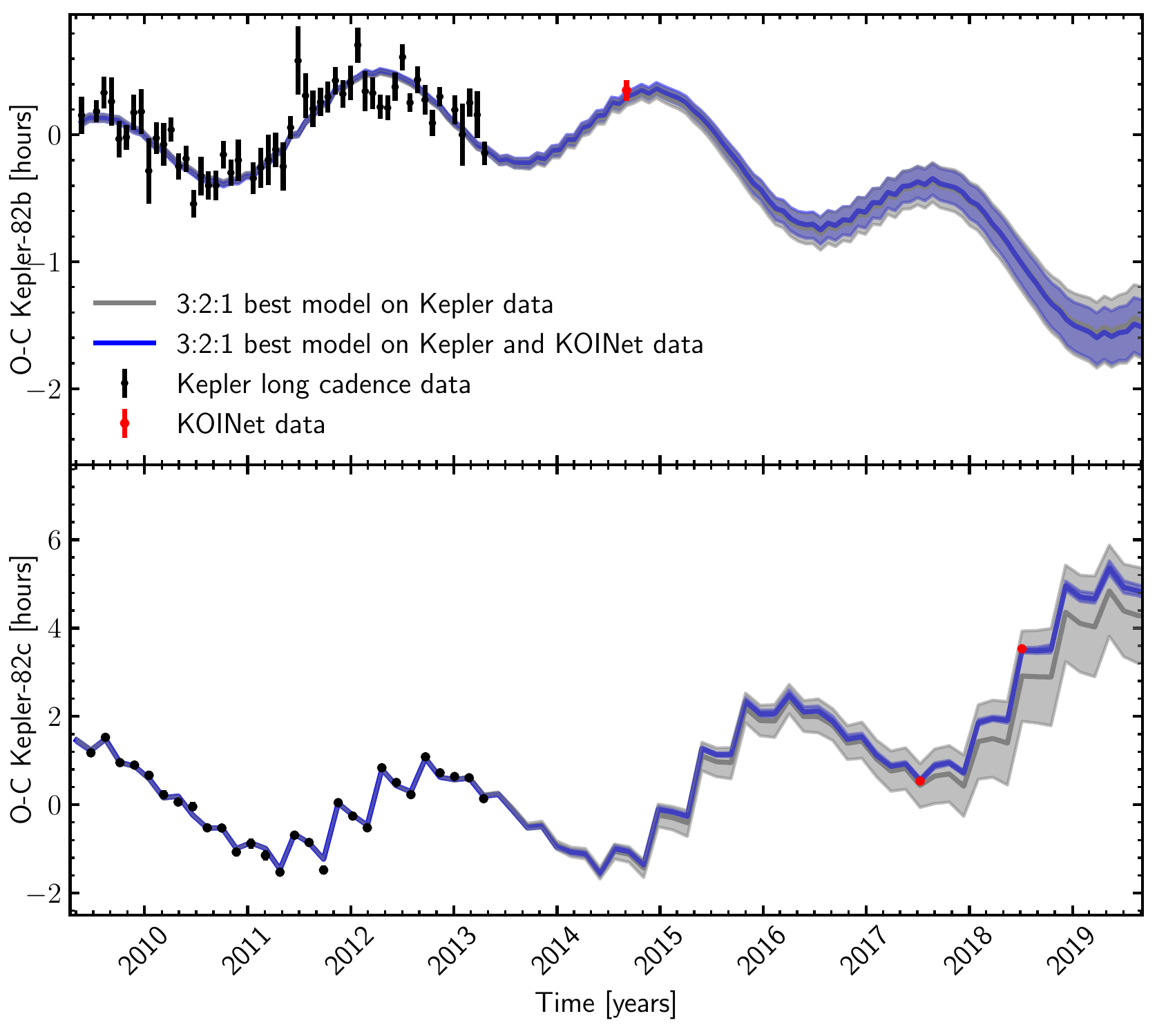}
 \caption{O-C diagram of Kepler-82b at the top and Kepler-82c at the bottom with transit times from modelling the transits individually. The black points refer to the transit data from the Kepler telescope. The red points are the individual transit times from the new KOINet observations. The grey area indicates the 68.3\;\% confidence interval of the 3:2:1 resonance model fitted to the Kepler long- and short-cadence data only, the grey line is the best $\chi^2$ solution. The blue area indicates the same model solution applied to \Kepler and KOINet data respectively.}
 \label{fig:ttv2_1}
\end{figure*}

\begin{figure*}
 \resizebox{\hsize}{!}{\includegraphics{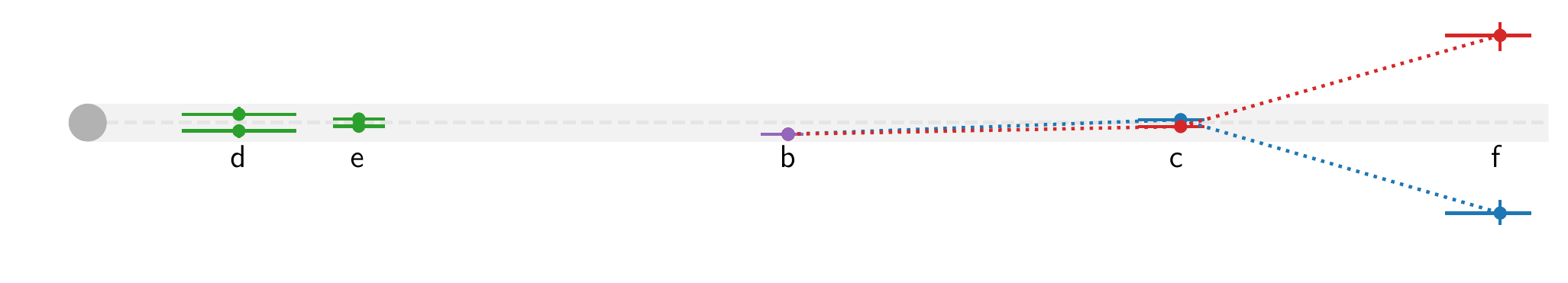}}
 \caption{The configurations of the Kepler-82 system. With the star in grey on the left side and the observer on the right side this shows the two different configurations: b in violet has the same position in both, c and f in red shows configuration~I and in blue configuration~II. The grey area indicates the region of impact parameters below one. The distances are true to scale with the stellar radius. Additionally, the two inner planets are plotted in green, the data are taken from the NASA~Exoplanet~Archive. They are plotted on both sides because we did not include them in the photodynamical analysis and hence we do not know how they behave in the two configurations. }
 \label{fig:config2}
\end{figure*}

\begin{figure*}
 \resizebox{\hsize}{!}{\includegraphics{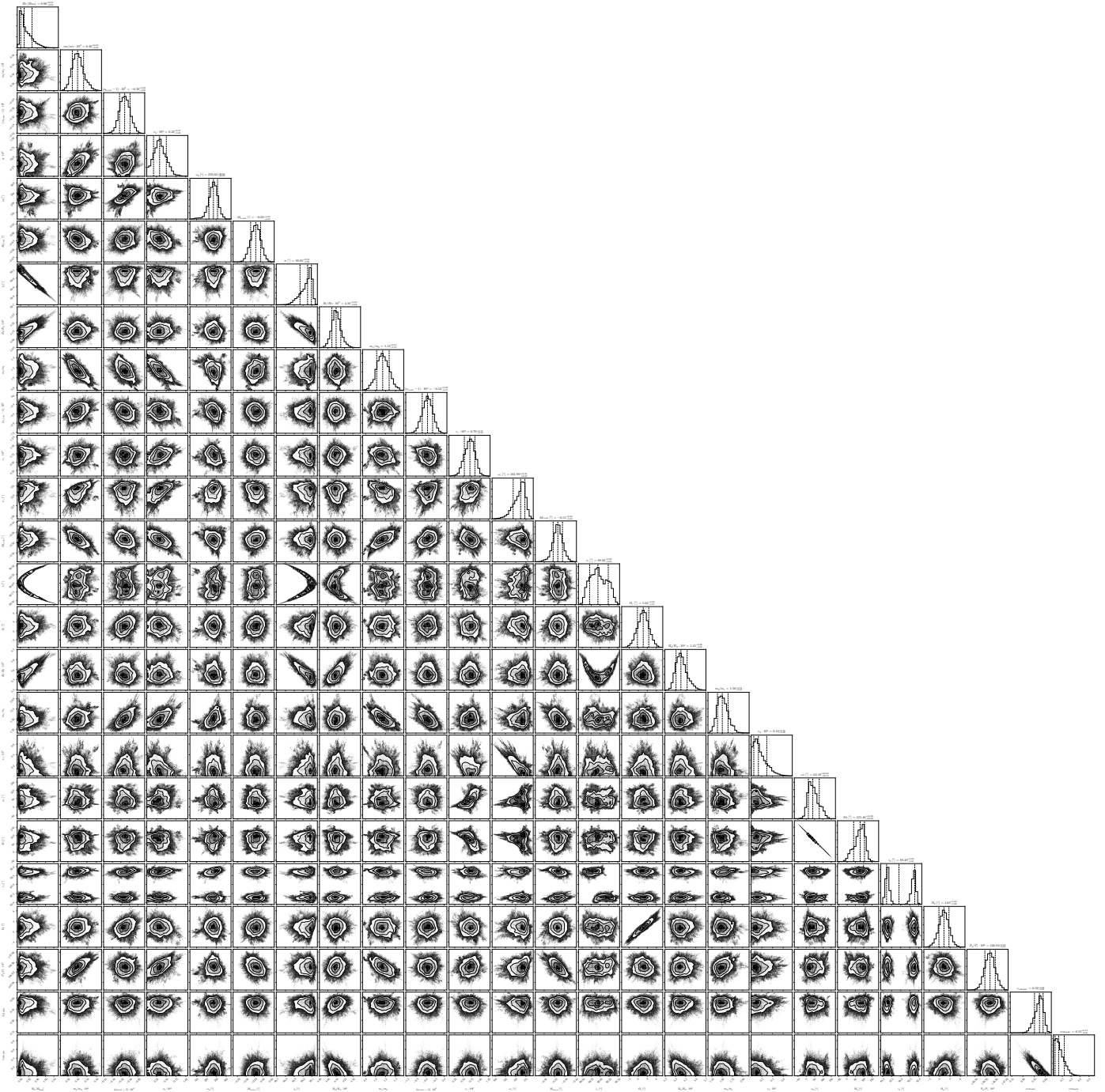}}
 \caption{Correlation plot of all fit parameters from modelling the 3:2:1 resonance model to \textit{Kepler} long- and short-cadence and KOINet data.}
 \label{fig:corner}
\end{figure*}

\end{appendix}
\end{document}